\documentclass[%
superscriptaddress,
reprint,
amsmath,amssymb,
aps,
prb,
]{revtex4-1}
\usepackage{graphicx}% Include figure files
\usepackage{dcolumn}% Align table columns on decimal point
\usepackage{bm}% bold math
\usepackage{hyperref}% add hypertext capabilities
\begin{document}

%\preprint{APS/123-QED}

\title{Exceptional points and their coalescence of $\mathcal{PT}$-symmetric interface states in photonic crystals}% Force line breaks with \\

\author{Xiaohan Cui}
\affiliation{%
	Department of Physics, Hong Kong University of Science and Technology, Clear Water Bay, Kowloon, Hong Kong, China
}%
\author{Kun Ding}
\affiliation{%
	Department of Physics, Hong Kong University of Science and Technology, Clear Water Bay, Kowloon, Hong Kong, China
}%

\affiliation{The Blackett Laboratory, Department of Physics, Imperial College London, London SW7 2AZ, United Kingdom}
\author{Jian-Wen Dong}
%\email{dongjwen@mail.sysu.edu.cn}
\affiliation{
School of Physics $ \& $ State Key Laboratory of Optoelectronic Materials and Technologies, Sun Yat-Sen University, Guangzhou 510275, China}
\author{C. T. Chan}%
 \email{phchan@ust.hk}
\affiliation{%
Department of Physics, Hong Kong University of Science and Technology, Clear Water Bay, Kowloon, Hong Kong, China
}%



\date{\today}% It is always \today, today,

\begin{abstract}
The existence of surface electromagnetic waves in the dielectric-metal interface is due to the sign change of real parts of permittivity across the interface. In this work, we demonstrate that the interface constructed by two semi-infinite photonic crystals with different signs of the imaginary parts of permittivity also supports surface electromagnetic eigenmodes with real eigenfrequencies, protected by $\mathcal{PT}$ symmetry of the loss-gain interface. Using a multiple scattering method and full wave numerical methods, we show that the dispersion of such interface states exhibit unusual features such as zigzag trajectories or closed-loops. To quantify the dispersion, we establish a non-Hermitian Hamiltonian model that can account for the zigzag and closed-loop behavior for arbitrary Bloch momentums. The properties of the interface states near the Brillouin zone center can also be explained within the framework of effective medium theory. It is shown that turning points of the dispersion are exceptional points (EPs), which are characteristic features of non-Hermitian systems. When the permittivity of photonic crystal changes, these EPs can coalesce into higher order EPs or anisotropic EPs. These interface modes hence exhibit and exemplify many complex phenomena related to exceptional point physics.

\end{abstract}

\maketitle

%\tableofcontents

\section{\label{sec:introduce}Introduction}

Interface states commonly exist in quantum systems and classical wave systems. A well-known example is the surface plasmon polaritons \cite{Barnes2003,Popov2008}, which are surface waves traveling along a dielectric-metal interface due to a change in sign of the real part of permittivity across the interface. Photonic crystal (PC) systems also have surface modes, and in some cases, the existence of the boundary modes can be explained using topological concepts \cite{Wang2009, Feng2011, Fang2012, Rechtsman2013,Huangxueqin2014, Xiao2016, Lu2016}. These prior topological-based research on interface states in photonic crystal systems were mainly focused on Hermitian systems. However, recent works show that boundary modes can also be found in non-Hermitian systems \cite{Longhi2014,Savoia2015,Weimann2016,Xuyadong2017, Shen2018,Yao2018,MartinezAlvarez2018, Ni2018}. For example, surface states can be localized at the gain-loss interface in $\mathcal{PT}$-symmetric  systems\cite{Longhi2014,Savoia2015}.

In this work, we study the formation of interface states in a non-Hermitian PC with a \(PT\)-symmetric interface, in which one side of the PC has gain and the other side has loss. We find that such a system carries interface states with real eigenvalues. The dispersions of these interface states are rather unusual, as they form ziz-zag trajectories or closed-loops. Exceptional points \cite{Bender1998,Heiss2012, Hodaei2014,Zhen2015,Hodaei2017,El-Ganainy2019,Miri2019}, which are characteristic features in non-Hermitian systems, appear at the turning points of the dispersions of the interface states. 
EPs are branch point singularities in parameter spaces, at which eigenvalues and eigenvectors coalesce simultaneously. At EPs, the matrix Hamiltonian is defective, and the coalescing eigenvectors are not linearly independent\cite{Rotter2009}, which is different from degenerate points in Hermitian systems.  
As the system parameters changes, we found that the EPs can coalesce into higher order EPs\cite{Graefe2008,Demange2012,Ding2015coalesce,Ding2016,Xulin2019} or anisotropic EPs \cite{Ding2018,Xunlin2018,Yixin2019}. Besides, we find that in the limit of large gain/loss, one band of interface states with real eigenvalues always persists.

We study the interface states using two different computation approaches (numerical package COSMOL and a multiple scattering method) and two different boundary conditions (periodic and open). These computation details are described in Sec. \ref{sec:2} and Sec. \ref{sec:3}. In Sec. \ref{sec:4}, we attempted to give a simple explanation to the rather exotic dispersions using effective medium theory (EMT), which works well near the Brillouin zone center. In Sec. \ref{sec:5}, we formulate a non-Hermitian Hamiltonian model for the interface states that works for a general value of the Bloch momentum and the model shows clearly that there are EPs in the band of interface states. In Sec. \ref{sec:6}, we show that EPs can coalesce into higher order EPs and anisotropic EPs as system parameters change. In the last section, we give a conclusion. 

\begin{figure*}
	\centering
	\includegraphics[width=0.75\linewidth]{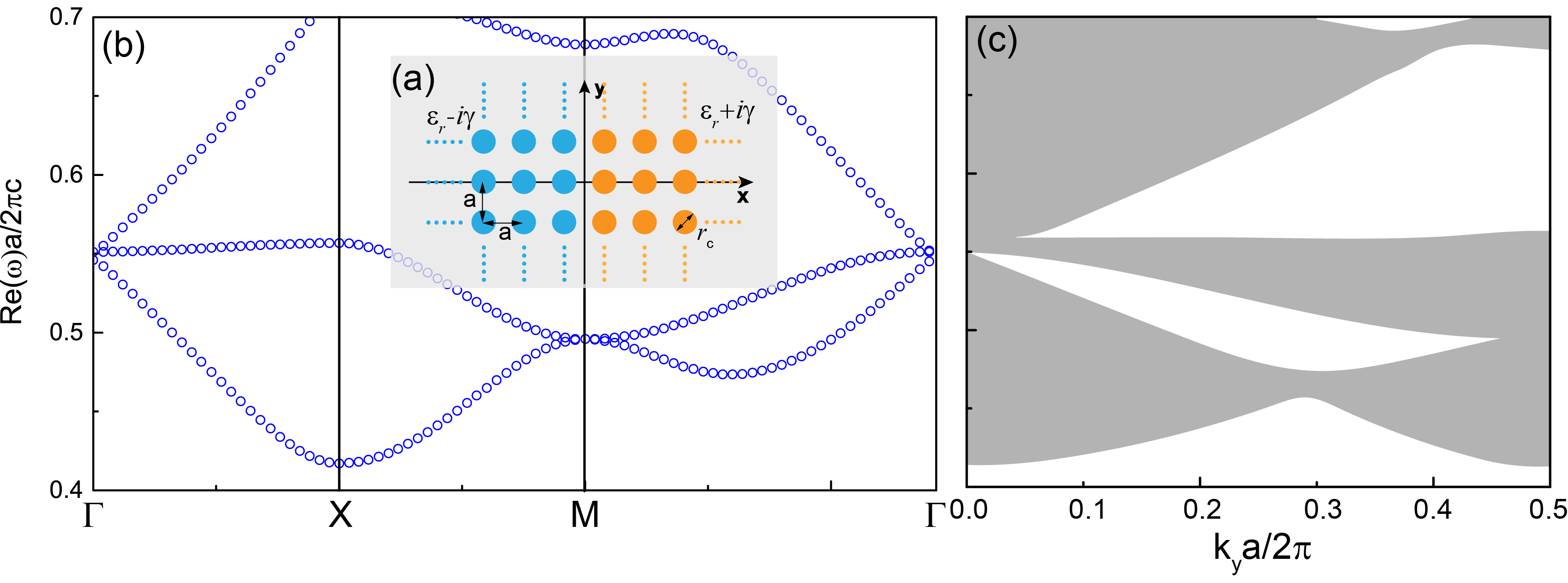}
	\caption{(Color online) (a) Schematic picture of an interface structure constructed by two semi-infinite 2D PCs consisting of dielectric cylinders with a radius of \(r_c\) and a relative permittivity $\varepsilon_c=\varepsilon_r\pm i \gamma$  embedded in air with a relative permittivity  $\varepsilon_b=1.0$, and the lattice constant is $a$. The relative permeability $\mu$ of both media is 1.0. (b) The band structures of a Hermitian (\(\gamma=0\)) 2D PC with parameters \( r_c=0.2a\), \(\varepsilon_c=12\) along $\Gamma-X-M-\Gamma$ directions. (c) The projected band structures along the \(k_y\) direction.}
	\label{fig:interface_band_diagram}
\end{figure*}

\section{\label{sec:2}PT-symmetric interface states in photonic crystals}
We consider a PC comprising of two different two-dimensional (2D) square lattice PCs, as illustrated in Fig. \ref{fig:interface_band_diagram}(a). In this work, we focus on the transverse magnetic (TM) polarization ( $E_z$ polarization).  Within the primitive unit cell with a lattice constant $a$, the rod has a radius of $r_c$ and a relative permittivity $\varepsilon_c=\varepsilon_r\pm i \gamma$  embedded in air. The relative permeability $\mu$  is 1.0 everywhere. The positive (negative) sign of $\gamma$ indicates the rod is a lossy (gain) medium. There is a \(PT\)-symmetric interface at \(x=0\) separating cylinders with gain on one side and lossy cylinders on the other side.
The cylinders (blue) of the left semi-infinite PC  (\(x<0\)) are active with gain ($\varepsilon_c=\varepsilon_r-i \gamma$) while the cylinders (orange) at the right semi-infinite PC (\(x>0\)) are lossy ($\varepsilon_c=\varepsilon_r+i \gamma$). 

\begin{figure*}
	\centering
	\includegraphics[width=0.75\linewidth]{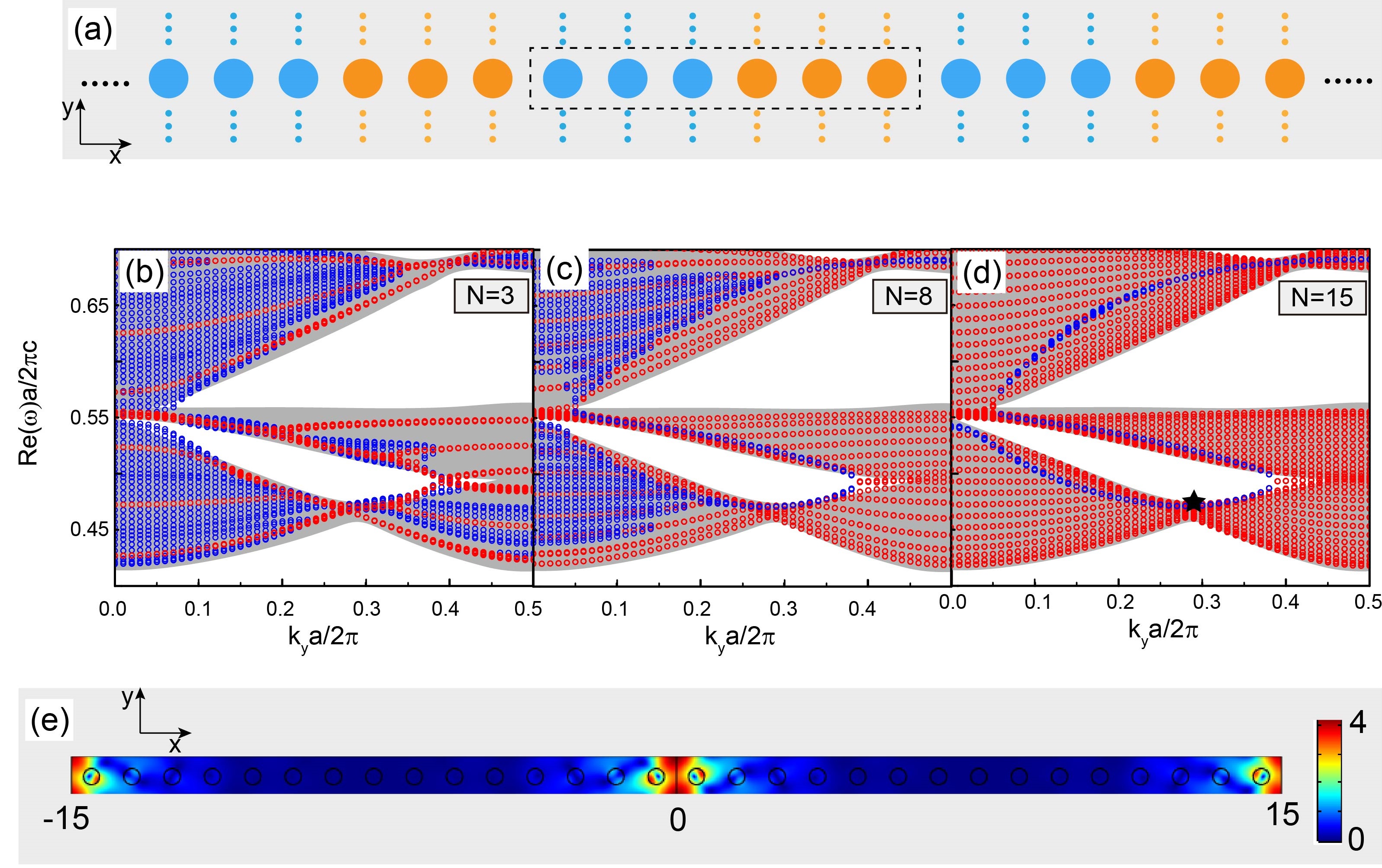}
	\caption{(Color online) (a) Schematic picture of PC of which the unit cell (dashed rectangle) has \(N=3\) active cylinders (orange) and \(N=3\) lossy cylinders (blue). Periodic conditions are applied to \(x\)- and \(y\)-direction. The projected bands of PC with parameters $ r_c=0.2a $ and $ \varepsilon_c=12\pm i $ along \(k_y\) direction are calculated using COMSOL and the numbers of cylinders in one unit cell  are (b) \(N=3\), (c) \(N=8\), (d) \(N=15\), respectively. The gray shadow region marks the projected bands of Hermitian case (\(\gamma=0\)). The real parts of the eigenfrequencies for states in $\mathcal{PT}$ exact phase (${\mathop{\rm Im}\nolimits} \left( \omega  \right)a/2\pi c{\rm{ = 0}}$) and $\mathcal{PT}$ broken phase (${\mathop{\rm Im}\nolimits} \left( \omega  \right)a/2\pi c \ne 0$ ) are plotted by blue and red circles, respectively. (e) The electric field distribution \(|\textbf{E}|\) of an interface state (labeled by a black star (d)) at the frequency $ Re (\omega) a/2 \pi c=0.47 $ and \( k_y a/2\pi=0.3\), \(k_x a/2\pi=0\). }
	\label{fig:interface_periodic_x_projectband}
\end{figure*}

At the Hermitian limit $ \gamma = 0 $, the $\mathcal{PT}$-symmetric interface disappears, and this system becomes a simple Hermitian 2D PC with the same material parameters everywhere. The bulk band structure and projected band diagrams along the \(k_y\) direction of the Hermitian PC with parameters \(r_c=0.2a,\varepsilon_c=12 \) are calculated using COMSOL as shown in Figs. \ref{fig:interface_band_diagram}(b) and \ref{fig:interface_band_diagram}(c), respectively. It has been shown theoretically and demonstrated experimentally that in  2D Hermitian PCs possessing bands with a Dirac-like cone dispersion at $k = 0$, even a small perturbation of the relative permittivity \(\varepsilon_r\) is sufficient to create interface states \cite{ XiaoPRX2014, Huangxueqin2014,Xiao2016,Hangzhihong2016}. The underlying physics for the existence of an interface state is related to the geometric phase of the bulk bands. In this work, we want to study the emergence of interface states upon the introduction of an imaginary part to the relative permittivity. More specifically, we create an interface by introducing \(PT\)-symmetric non-Hermiticity into this Hermitian 2D PCs as depicted in Fig. \ref{fig:interface_band_diagram}(a).

In numerical calculations, the number of column layers of the semi-infinite PC is truncated to a finite number \(N\). One method of terminating the far ends of the semi-infinite PC is to apply open boundary conditions, such as perfectly matched layers (PML) or scattering boundary along \(x\)-direction. Thus, the \(PT\)-symmetric PC is periodic along \(y\)-direction but is finite along \(x\)-direction. Another method to study the interface states is to apply periodic boundary condition also along \(x\)-direction. Figure \ref{fig:interface_periodic_x_projectband}(a) gives a schematic picture of such a lattice, with a unit cell containing \(N=3\) cylinders with gain and \(N=3\)  cylinders with loss. Both boundary conditions give the same result in the limit of large \(N\), showing that in this particular case the results are not dependent on boundary conditions even though the system is non-Hermitian \cite{Lee2016}. The periodic condition gives us a more heuristic understanding on the formation of the interface bands from the point view of \(PT\)-symmetry, which has been extensively studied in prior research.

We first calculate the band structure of the \(PT\)-symmetric PC with periodic boundary conditions. As shown in Fig. \ref{fig:interface_periodic_x_projectband}(a), the unit cell (marked by black dashed rectangle) of the lattice contains \(N\) active cylinders (orange) together with \(N\) lossy cylinders (blue). The gray shadow region of Figs.  \ref{fig:interface_periodic_x_projectband}(b-d) marks the projected bands of Hermitian PC with parameters \( r_c=0.2a\), \(\varepsilon_c=12\), which is the same as the projected band shown in Fig. \ref{fig:interface_band_diagram}(c). We then calculate the projected bands of \(PT\)-symmetric PC with parameters \(\varepsilon_c=12 \pm i\) using COMSOL. In Fig. \ref{fig:interface_periodic_x_projectband}(b-d), we plot the real parts of eigenfrequencies of PC, and the number \(N\) of gain/loss cylinders in the unit cell is labeled in the figure. The imaginary parts of the eigenfrequency of eigenstates marked by blue circles are zero, and those by red are nonzero. For \(PT\)-symmetric systems, the eigenstates with purely real eigenvalues are said to be in the exact \(PT\) symmetry phase, and those with complex-conjugate-pairs eigenvalues are said to be in the broken \(PT\) symmetry phase \cite{Bender2007}. From Fig. \ref{fig:interface_periodic_x_projectband}(b), we see that when we introduce the \(PT\)-symmetric non-Hermiticity into Hermitian supercells of the 2D PC, some eigenstates (marked by red circles) acquire imaginary parts and go into the broken \(PT\)-symmetry phase. When the supercell is small (e.g. \(N=3\)), most of the eigenstates (marked by blue circles) are purely real, and these eigenstates are still in the exact \(PT\) symmetry phase. When we increase the number of cylinders in the unit cell \( N\) to 8, an increasing number of the eigenstates go into the broken \(PT\) symmetry phase as depicted in Fig. \ref{fig:interface_periodic_x_projectband}(c). When the number of cylinders in the unit cell \(N=15\) as shown in Fig. \ref{fig:interface_periodic_x_projectband}(d), nearly all of the eigenstates in the whole Brillouin zone are in the broken \(PT\) symmetry phase, except for one band of states with a zigzag dispersion which remains in the exact \(PT\) symmetry phase. In Fig. \ref{fig:interface_periodic_x_projectband}(e), we plot one of these states in \(PT\) symmetry exact phase (marked by dark star in Fig. \ref{fig:interface_periodic_x_projectband} (d)), and find that the electric field distribution \(|\textbf{E}|\) is localized at the interface \(x=\{-15a,0,15a\}\) separating the gain/loss cylinders, i.e., the \(PT\)-symmetric interface. 
The interface modes lie inside the continuum of the real-valued bulk modes when $ \gamma=0 $ (gray shadow region), but when $ \gamma \ne0 $, the continuum becomes complex-valued. Therefore, in a certain sense these interface modes are bound states in the continuum with complex-valued energy spectra. This is different from the conventional “bound states in continuum”, where the continuum is real-valued\cite{Longhi2014,Longhi2014BIC}.

We emphasize that in going from panels (b) to (d) of Fig. \ref{fig:interface_periodic_x_projectband}, the material parameters and in particular the imaginary parts of the permittivity remain unchanged. The only change is the number of cylinders in the unit cell. There are both localized interface states and extended bulk states in this $\mathcal{PT}$-symmetric PC. For a fixed $ \gamma=1 $, as $N$ increases, the bulk states undergo a phase transition from the exact \(PT\) regime to the broken \(PT\) regime. For bulk states at different Bloch $ k_y $, the threshold value of \(N\) for the phase transition point is different. When the bulk states are in the broken \(PT\) regime, the field distribution is either concentrated in the lossy medium or in the gain medium. If the bulk state eigenfields are mainly concentrated in the left gain domain, the eigenfrequencies have negative imaginary parts. If the bulk states are mainly localized in the right lossy domain, the eigenfrequencies have positive imaginary parts. However, the fields of interface states are always localized at the loss-gain interface, and the field distributions are symmetric about the interface even for large value of $N$. Accordingly, the interface states persist in the exact \(PT\) regime and have purely real eigenfrequencies. By studying how the eigenstates change as the number of cylinders $ N $ in the unit cell increases, we distinguish the localized interface states and extended bulk states based on whether their eigenvalues are real numbers.

\section{\label{sec:3} $\mathcal{PT}$-symmetric interface states using multiple scattering method with open boundary condition}

When the column number \(N\) is large enough, the results calculated with periodic condition should be the same with open boundary conditions. In this section, we calculate the band structure using a multiple scattering (MS) method with open boundary conditions, and we are expected to observe the interface states plotted in Fig. \ref{fig:interface_periodic_x_projectband}(d) calculated using periodic boundary conditions. We use a Green's function method to build a scattering problem of cylinders arranged in a square lattice grid, the details of which are given in Appendix \ref{sec:appA}. In the TM polarization (the electric field is along the axes of cylinders), the dominant excitations of the cylinders at low frequencies are the out-of-plane electric monopole and two in-plane magnetic dipoles. The scattering problem for plane waves incident on the PC can be written as below,
\begin{equation}\label{eq:c5-scatteringproblem}
M(\omega, k_y) \Psi^{\mathrm{loc}}=\Psi^{\mathrm{inc}},
\end{equation}
where $ \Psi^{\mathrm{inc}} $ represents external incident waves and $ \Psi^{\mathrm{loc}} $ represents total local fields. The band structure of a PC can be obtained by solving an eigenvalue problem in the absence of incident field,
\begin{equation}\label{eq:c5-scatteringeigenvalueproblem}
M(\omega, k_y) \Psi^{\mathrm{loc}}=0.
\end{equation}
Frequencies $ \omega $ for which Eq. \eqref{eq:c5-scatteringeigenvalueproblem} has non-trivial solutions for a given Bloch $ k_y $ are zeros of matrix $ M(\omega, k_y) $, which can be found by locating the zeros of its determinant,
\begin{equation}
\operatorname{det}\left[M\left(\omega, k_{y}\right)\right]=0.
\end{equation}
As the calculation of the determinant becomes numerically unstable if the matrix is large, we adopt the numerically stable approach to solve the equation
\begin{equation} \label{eq:c5-minS}
    {\rm{Min}}\left\{ {\left| {{\mathop{\rm eig}\nolimits} \left[ {{M}\left( {\omega ,{k_y}} \right)} \right]} \right|} \right\} = 0,
\end{equation}
where $ {\rm{Min}}\left\{ {\left| {{\mathop{\rm eig}\nolimits} \left[ {{M}\left( {\omega ,{k_y}} \right)} \right]} \right|} \right\} $ is the minimal-modulus eigenvalue of the matrix $ M\left(\omega, k_{y}\right) $.
Therefore, equation \eqref{eq:c5-minS} defines the dispersion curves $ \omega (k_y) $ for photons propagating through the periodic structure. The MS method is very useful as it can handle non-Hermitian systems and is compatible with various types of boundary conditions.

\begin{figure}
    \centering
    \includegraphics[width=\linewidth]{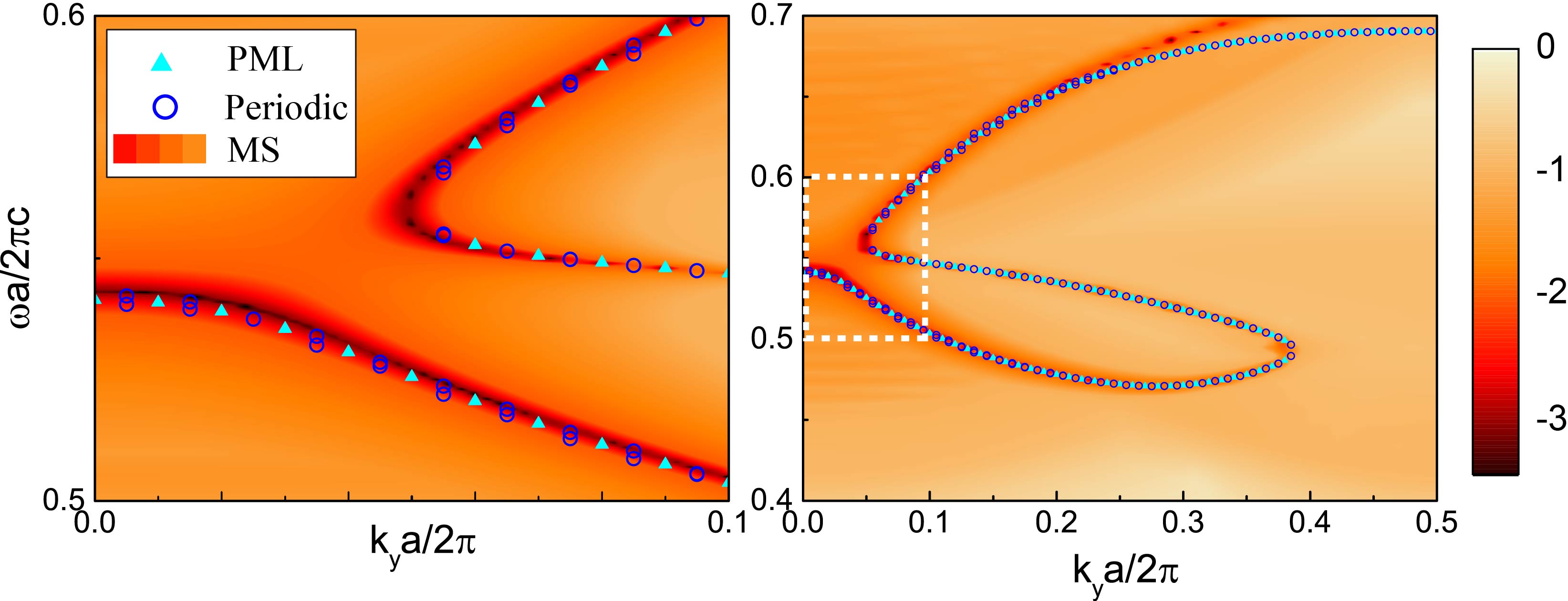}
    \caption{ (Color online) Colormap plot of \({\rm{Min}}\left\{ {\left| {{\mathop{\rm eig}\nolimits} \left[ {{M}\left( {\omega ,{k_y}} \right)} \right]} \right|} \right\}\) in log-10 scales as functions of real frequency $\omega$ and real Bloch wave vector $k_y$ for non-Hermitian PC with parameters $\varepsilon_c=12 \pm i$. The number of column layers of the PC is truncated to $N=15$. Interface states calculated by COMSOL with periodic and PML boundary conditions applied on the $ x $-direction are plotted by blue circles and cyan triangles, respectively. The left panel is the enlargement of the white dashed rectangle in the right panel.}
    \label{fig:PMLvsPeriodicvsMS}
\end{figure}

Using the MS method, we calculate the interface states of a $\mathcal{PT}$-symmetric PC with $\varepsilon_c=12 \pm i$. The column number of the unit cell of the PC is truncated to $N=15$, and scattering boundary conditions applied in the \(x\)-direction . In Fig. \ref{fig:PMLvsPeriodicvsMS}, we plot the values of \({\rm{Min}}\left\{ {\left| {{\mathop{\rm eig}\nolimits} \left[ {{M}\left( {\omega ,{k_y}} \right)} \right]} \right|} \right\}\) in log-10 scale, in the parameter plane $\left( \omega, k_y \right)$, where $\omega$ and $k_y$ are both real numbers. The value \({\rm{Min}}\left\{ {\left| {{\mathop{\rm eig}\nolimits} \left[ {{M}\left( {\omega ,{k_y}} \right)} \right]} \right|} \right\}\) of the states with complex eigenfrequencies cannot be zero at real-valued $\left( \omega, k_y \right)$ plane.
Therefore, the dark orange lines, marking locations of \({\rm{Min}}\left\{ {\left| {{\mathop{\rm eig}\nolimits} \left[ {{M}\left( {\omega ,{k_y}} \right)} \right]} \right|} \right\}\to 0\) represent interface eigenstates with real eigenfrequencies. In the vicinity, the areas marked by lighter colors (\({\rm{Min}}\left\{ {\left| {{\mathop{\rm eig}\nolimits} \left[ {{M}\left( {\omega ,{k_y}} \right)} \right]} \right|} \right\}>0\)) indicate the bulk states with complex eigenfrequencies.
Using the MS method, we can locate the interface states with purely real eigenfrequencies.

To verify the interface states obtained by MS method, we also calculated the eigenstates by COMSOL, with PML boundary conditions applied to the \(x\)-direction, whereas a periodic boundary condition is applied to the \(y\)-direction for each wave number \( k_y\). Using COMSOL, we pick the eigenstates with real eigenvalues and plot them in Fig. \ref{fig:PMLvsPeriodicvsMS} by cyan triangles. We also plot the interface states calculated by COMSOL with periodic conditions applied to the \(x\)-directions by blue circles in Fig. \ref{fig:interface_periodic_x_projectband}(d) in the same picture.  We can see from Fig. \ref{fig:PMLvsPeriodicvsMS} that the results calculated by COMSOL using PML and periodic boundary conditions agree well with the results calculated by the MS method.

\section{\label{sec:4}Effective Medium Theory}

\begin{figure}
	\centering
	\includegraphics[width=\linewidth]{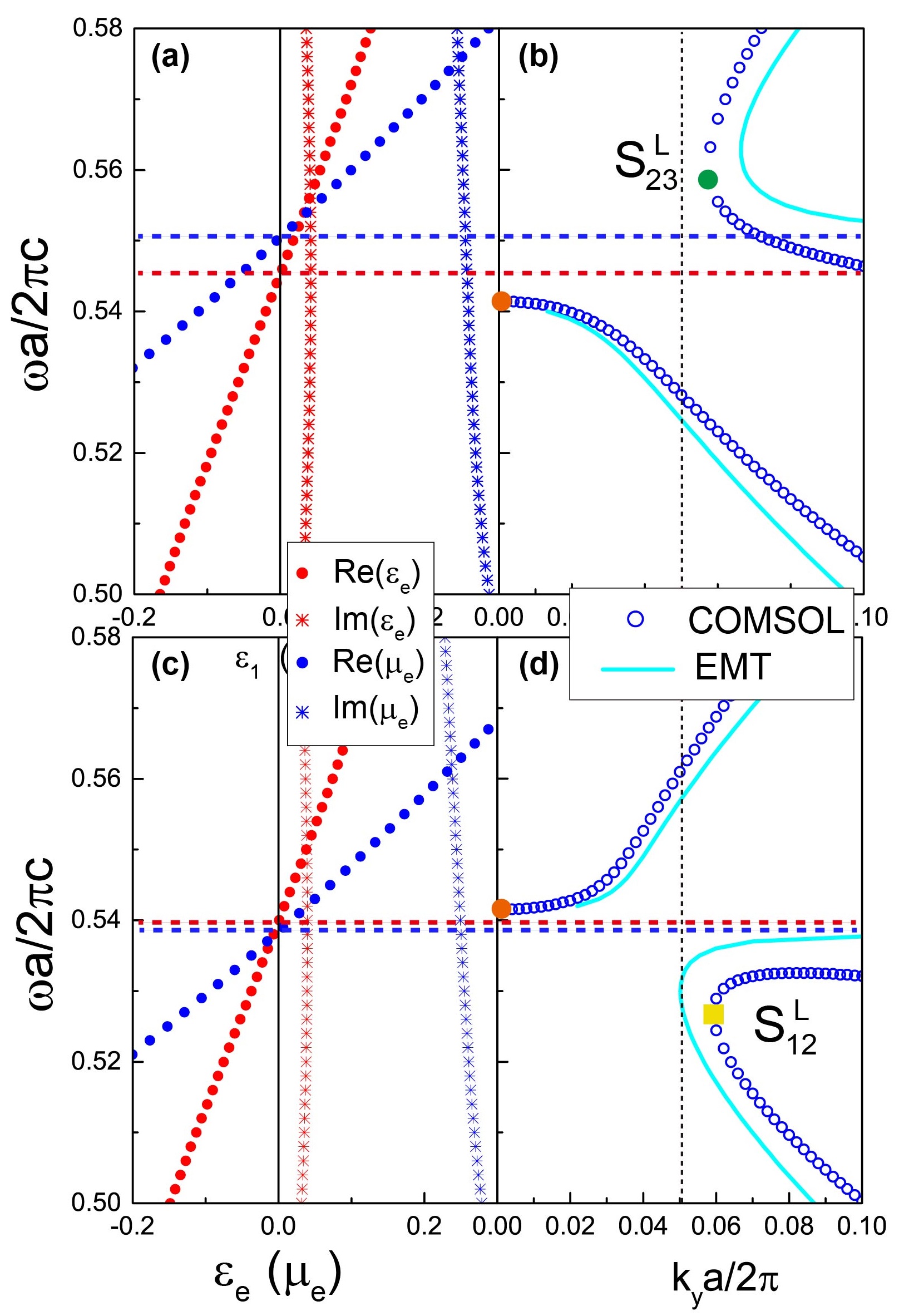}
	\caption{(Color online) Effective parameters $ \varepsilon_e$, $\mu_e $ of non-Hermitian PC with parameters (a) $\varepsilon_c=12+1.2i$ and (c) $\varepsilon_c=12.6+1.2i$ are plotted in left panels. The band structure of interface states of $\mathcal{PT}$-symmetric PC with (b) $\varepsilon_c=12 \pm 1.2i$ and (d) $\varepsilon_c=12.6 \pm 1.2i$ are plotted in right panels. The blue open circles in (b), (d) are calculated by COMSOL with PML boundary conditions and the column number of the semi-infinite PCs is truncated to $N=15$. The cyan lines denote the dispersion (Eq. \eqref{eq:interface_EMT_dispersion}) of interface states of $\mathcal{PT}$-symmetric homogeneous slab with $ \varepsilon_e$, $\mu_e $. The frequencies $ \omega_{ e} $, $ \omega_m $ and $ \omega_\beta $ are labeled by red dashed lines, blue dashed lines and orange dots, respectively. Two EPs $ S_{23}^L $ and $ S_{12}^L $ are marked by green dot and yellow square, respectively. The vertical black dashed line labels ${k_y}a/2\pi  = 0.05$. The radius of the cylinders of the PC is $r_c=0.2a$.}
	\label{fig:EMTandBandinversion}
\end{figure}

%\cite{*[{}] [{. This work under review.}] CCM}.
We now consider the interface states from the viewpoint of effective medium theory (EMT). We will see whether an effective medium description in the long wavelength limit can provide a simple heuristic picture to understand the formation of interface states near the zone center. The effective parameters of non-Hermitian PCs can be obtained conveniently using a boundary field averaging method \cite{Andryieuski2012}. To obtain the effective parameters of the PC, we need to treat the PC as a homogeneous medium. Assuming that a plane wave propagates along the \(x\)-direction with wave vector \(k \hat x\) and polarization \(E_z  \hat z\) in a homogeneous medium, we define a wave impedance,
\begin{equation}\label{eq:waveimpedance}
Z=\frac{E_{z}}{H_{y}}=-\frac{\omega \mu}{k}=-\frac{k}{\omega \varepsilon}
\end{equation}
For the inhomogeneous PC system with a micro-structure, we define a corresponding average field ratio as,
\begin{equation}
Z_B=\frac{\int_{I} E_{z}^{\mathrm{PC}} d y}{\int_{I} H_{y}^{\mathrm{PC}} d y},
\end{equation}
where $E_{z}^{\mathrm{PC}}$ and $H_{y}^{\mathrm{PC}}$ are the eigenfield at the incident boundary $I$ of the unit cell. When using EMT, the frequency $ \omega $ in Eq. \eqref{eq:waveimpedance} should take real number, because the effective parameters are functions of real frequency. To obtain eigenfields $E_{z}^{\mathrm{PC}}$ and $H_{y}^{\mathrm{PC}}$, we use COMSOL to solve a complex-valued \(\textbf{k}(\omega)\) vs. real-valued \(\omega\) dispersion \cite{Davanco2007,Fietz2011}.  According to Eq. \eqref{eq:waveimpedance}, the effective permittivity and permeability are
\begin{equation}\label{eq:effectiveepsilonandmur}
\varepsilon_{\mathrm{e}}=-\frac{k}{\omega \varepsilon_{0} Z_B}, \qquad \mu_{\mathrm{e}}=-\frac{k}{\mu_{0} \omega} Z_B.
\end{equation}
We first calculate the effective parameters $\varepsilon_{\rm e}$ and $\mu_{\rm e}$ for the right lossy semi-infinite PC with $\varepsilon_c=\varepsilon_r+i \gamma$. Therefore, the effective parameters for the left active semi-infinite PC with $\varepsilon_c=\varepsilon_r-i \gamma$ are $\varepsilon_{\rm e}^*$ and $\mu_{\rm e}^*$. From the point of view of EMT, the PT-symmetric PC illustrated in Fig. \ref{fig:interface_band_diagram}(a) can be treated as a $\mathcal{PT}$-symmetric homogenous slab with permittivity and permeability like below,
\begin{equation}\label{eq:effectivePT}
\varepsilon(x)=\left\{\begin{array}{lll}{\varepsilon_{\mathrm e}^{*},} & {x<0} \\ {\varepsilon_{\mathrm e},} & {x>0}\end{array}, \quad \mu(x)=\left\{\begin{array}{ll}{\mu_{\mathrm e}^{*},} & {x<0} \\ {\mu_{\mathrm e},} & {x>0}\end{array}\right.\right. .
\end{equation}
Enforcing the continuity condition at the interface \( x=0\) for the electric field, a modal solution bound at the gain-loss interface can be written as \cite{Savoia2015}
\begin{equation}
E_{z}(x, y)=C \exp (i \beta y)\left\{\begin{array}{ll}{\exp \left(i k_{x}^{*} x\right),} & {x<0} \\ {\exp \left(i k_{x} x\right),} & {x>0}\end{array}\right.
\end{equation}
where \(C\) denotes a normalization constant, \(\beta \) is the propagation constant. We suppose the interface states are propagating along the \(y\)-direction without attenuation, which means that \(\beta\) is a purely real number, and therefore, we obtain that \(k_{x}^{2}=\varepsilon_{\mathrm{e}} \mu_{\mathrm{e}}(\omega / c)^{2}-\beta^{2}\) for \(x>0\) and  \(k_{x}^{*2}=\varepsilon_{\mathrm{e}}^* \mu_{\mathrm{e}}^*(\omega / c)^{2}-\beta^{2}\) for \(x<0\). $c$ is the speed of light in vacuum. From Maxwell’s equation, we then calculate the tangential magnetic field,
\begin{equation}
	H_{y}(x, y)=\frac{-k_{x}(x)}{\omega \mu(x)} E_{z}(x, y).
\end{equation}
By enforcing its continuity at the interface, we obtain\cite{Lawrence2010}
\begin{equation}
	\frac{k_{x}}{\mu_{\mathrm{ e}}}=\frac{k_{x}^{*}}{\mu_{\mathrm{e}}^{*}}.
\end{equation}
 To obtain states bounded at $\mathcal{PT}$-symmetric interface, the complex number $k_x$ should satisfy $\operatorname{Im}\left(k_{x }\right)>0$ to ensure that the electric field exponentially decays for $ x>0 $. Accordingly, we have $\operatorname{Im}\left(k_{x}^*\right)<0$ for $ x<0 $, which also ensures the electric field exponentially decays away from the interface. We then solve the dispersion relationship like below
\begin{equation}
    \frac{\mu_{\rm e}}{\mu_{\rm e}^{*}}=\frac{k_{x}}{k_{x}^{*}}=\frac{i \sqrt{\beta ^{2}-\varepsilon_{\rm e} \mu_{\rm e}(\omega/c)^{2}}}{-i \sqrt{\beta^{2}-\varepsilon_{\rm e}^{*} \mu_{\rm e}^{*} (\omega/c)^{2}}},
\label{eq:interface_EMT_equation}
\end{equation}
and obtain
\begin{equation}
  \beta=\frac{\omega}{c} \sqrt{\frac{\mu_{\rm e} \mu_{\rm e}^{*}\left(-\varepsilon_{\rm e} \mu_{\rm e}^{*}+\varepsilon_{\rm e}^{*} \mu_{\rm e}\right)}{\mu_{\rm e}^{2}-\left(\mu_{\rm e}^{*}\right)^{2}}}=\frac{\omega\left|\mu_{\rm e}\right|}{c} \sqrt{\frac{1}{2}\left(\frac{\varepsilon^{\prime}}{\mu^{\prime}}-\frac{\varepsilon^{\prime \prime}}{\mu^{\prime \prime}}\right)},
\label{eq:interface_EMT_dispersion}
\end{equation}
where we have set $\mu_{\rm e}=\mu^{\prime}+i \mu^{\prime \prime}, \varepsilon_{\rm e}=\varepsilon^{\prime}+i \varepsilon^{\prime \prime}$. 
 
Using Eq. \eqref{eq:effectiveepsilonandmur}, we calculate the effective parameters $\varepsilon_{\rm e}$ and $\mu_{\rm e}$ of the lossy PC with relative permittivity $\varepsilon_c=12+1.2i$ and $\varepsilon_c=12+1.6i$. The results are plotted in Figs. \ref{fig:EMTandBandinversion}(a) and \ref{fig:EMTandBandinversion}(c), respectively. The horizontal axes represent the real parts or imaginary parts of the effective parameters. Using the dispersion relation described by Eq. (\ref{eq:interface_EMT_dispersion}), we can calculate the band dispersion of the interface states for the $\mathcal{PT}$-symmetric homogenous slab with effective parameters described by Eq. \eqref{eq:effectivePT}. The dispersions are plotted in Figs. \ref{fig:EMTandBandinversion}(b) and \ref{fig:EMTandBandinversion}(d) by cyan lines. The propagation constant \(\beta\) in the effective homogeneous medium is the Bloch \(k_y\) in the PC. For comparison, we further calculate the band structure of interface states near the Brillouin zone center using COMSOL and plot the results in Figs. \ref{fig:EMTandBandinversion}(b) and \ref{fig:EMTandBandinversion}(d) by open blue circles, which agree reasonably well with the cyan lines. 

Since EMT is accurate in the long wavelength limit, where $k_y$ is a small number, we can give an analytical explanation to the band diagrams of the interface states near the Brillouin zone center. From Figs. \ref{fig:EMTandBandinversion}(a) and \ref{fig:EMTandBandinversion}(c), we see that $ \mu ' (\omega) $ goes from negative to positive as frequency increases, and it becomes zero at a particular frequency, which we call $ \omega_m $. In Fig. \ref{fig:EMTandBandinversion}, we denote the frequency \(\omega=\omega_m\) satisfying $ \mu '(\omega_m)=0 $ by blue dashed lines. It is shown by Eq. (\ref{eq:interface_EMT_dispersion}) that $ k_y \to \infty $  at frequency $ \omega=\omega_m $ due to  $ \mu '(\omega_m)=0 $, which is depicted by the cyan lines in Figs. \ref{fig:EMTandBandinversion}(b) and \ref{fig:EMTandBandinversion}(d). From Figs. \ref{fig:EMTandBandinversion}(a) and \ref{fig:EMTandBandinversion}(c), we can see that the imaginary parts (denoted by asterisks) of the effective parameters are nearly constant functions of frequency while the real parts are linear functions of frequency. Therefore, near the Brillouin zone center, we can write a reasonably good approximation that \(\varepsilon^{\prime}=p\left(\omega-\omega_{ e}\right), \quad \mu^{\prime}=q\left(\omega-\omega_{m}\right)\) and $ \varepsilon^{\prime \prime} / \mu^{\prime \prime}=\rho $. $ p$, $q$, and $\rho $ are all positive numbers, and $ \omega_e $ is the frequency satisfying $ \varepsilon'(\omega_e)=0 $. We label the frequency $ \omega=\omega_e $ by red dashed lines in Fig. \ref{fig:EMTandBandinversion}. Then Eq. (\ref{eq:interface_EMT_dispersion}) becomes
\begin{equation}
k_{y}(\omega)=\frac{\omega\left|\mu_{\rm e}\right|}{c} \sqrt{\frac{1}{2}\left(\frac{p}{q} \frac{\omega-\omega_{\rm e}}{\omega-\omega_{m}}-\rho\right)}
\label{eq:interface-kyomega}
\end{equation}
By setting $ k_y(\omega)=0 $, we can solve the frequency $ \omega_\beta $ satisfying $ k_y(\omega_{\beta})=0 $  as
\begin{equation}
\omega_{\beta}=\frac{\omega_{\rm e}\left[1-\eta\left(\omega_{m} / \omega_{\rm e}\right)\right]}{1-\eta},
\label{eq:interface-omegabeta}
\end{equation}
where $ \eta=\rho q / p \ll 1 $  in the considered frequency range.  In Fig. \ref{fig:EMTandBandinversion}, we label the frequency \(\omega_\beta\) satisfying \(k_{y}\left(\omega_{\beta}\right)=0\) by orange dots

From Eq. \ref{eq:interface-omegabeta}, we see that if $ \omega_m > \omega_e $, then we have $ \omega_e >\omega_{\beta} $ , which is the case illustrated in Figs. \ref{fig:EMTandBandinversion}(a) and \ref{fig:EMTandBandinversion}(b). On the other hand, if $ \omega_m < \omega_e $, we have  $ \omega_e <\omega_{\beta} $, which is the case illustrated in Figs. \ref{fig:EMTandBandinversion}(c) and \ref{fig:EMTandBandinversion}(d). 
The inversion of ordering ( $ \omega_m> \omega_\beta $ in Fig. \ref{fig:EMTandBandinversion}(b) while $ \omega_m<\omega_\beta $ in Fig. \ref{fig:EMTandBandinversion}(d)) leads to a drastic change in the dispersion. The frequency range between \(\omega_m\) and \(\omega_\beta\) is a band gap induced by quasi-longitudinal resonance. Hence, \(\omega_e\), \(\omega_m\) and \(\omega_\beta\), which are determined by \(\varepsilon_r\), govern the band structure’s pattern of the interface states. 

%Many literatures consider various folded bands, and our work show a reminiscent of those.

%This is very similar to the EMT of a Hermitian PC \cite{Wu2006}. By employing accidental degeneracy, 2D PC with parameters $ \varepsilon_r=12.5 $ and $ r_c=0.2 a $ can exhibit Dirac-like cone dispersion near the Brillouin zone center \cite{Huang2011}. At the Dirac-like point frequency ($\omega=\omega_{\rm e}=\omega_{\rm m} $), three bands are degenerate, and the PC behaves like a zero-index medium with effective parameters $ \varepsilon_e=\mu_e=0 $. If we tune $ \varepsilon_r<12.5 $, the higher two bands are degenerate, and the relation of effective parameters become $ \omega_{ e}<\omega_{m} $.  If we tune $ \varepsilon_r>12.5 $, the lower two bands are degenerate, and the relation of effective parameters become $ \omega_{ e}>\omega_{m} $.    

\section{\label{sec:5}Non-Hermitian Hamiltonian model and exceptional points of the interface states}

In the above section, we study the dispersion of interface states as a function of $k_y$. In this section, we will formulate a non-Hermitian Hamiltonian model of the interface states as functions of \( \varepsilon_c=\varepsilon_r\pm \gamma \) for a fixed $ k_y $ \cite{Ding2015coalesce,Cui2019}.
For a 2D PC of which the cylinders are uniform in the \(z\)-direction (\(\varepsilon(\textbf{r})\) is independent of \(z\) coordinate), we consider the TM polarization with electric fields only having \(z\) component $ E_z (\textbf {r} )$. The wave vector \(\textbf{k}\) is parallel to the 2D \(x-y\) plane, and the electromagnetic fields   $ E_z (\textbf {r} )$ , $ H_x (\textbf {r} )$ and $ H_y (\textbf {r} )$ are also independent of the \(z\) coordinate.  The eigenfrequency and eigenstate of this 2D PC with TM polarization can be obtained by solving the following equation \cite{sakoda2004optical}
\begin{equation} 
	\left[\nabla^{2}+\left(\frac{\omega}{c}\right)^{2} \varepsilon(\mathbf{r})\right] E_z(\mathbf{r})=0,
	\label{eq:inter-Helmo}
\end{equation}
where \(\nabla ^2=\partial^2 x+\partial^2 y\), and \(\textbf{r}\) denotes the 2D position vector (\(x,y\)).

To calculate the eigenfrequencies of the interface states of non-Hermitian PCs with material parameters \( \varepsilon(\textbf{r})\), we construct a model Hamiltonian using the Bloch states of the interface bands of a PC with material parameters \( \varepsilon^{(0)}(\textbf{r}) \) as the bases. Here, \( \varepsilon^{(0)}(\textbf{r}) \) is the relative permittivity of the original PC, and \( \varepsilon(\textbf{r})\) is the modified relative permittivity of the PC under consideration. 
The relative permittivity of the original $\mathcal{PT}$-symmetric PC in Fig. \ref{fig:interface_band_diagram}(a) can be described as
\begin {equation} \label{eq:c5-parameterepsilon}
\varepsilon^{(0)}(\mathbf{r})=\left\{\begin{array}{cccc}{\varepsilon_{r}^{(0)}-i \gamma^{(0)}} & {\left|\mathbf{r}-\mathbf{r}_{i j}\right|<r_{c}} & {\text { and }} & {x<0} \\ {\varepsilon_{r}^{(0)}+i \gamma^{(0)}} & {\left|\mathbf{r}-\mathbf{r}_{i j}\right|<r_{c}} & {\text { and }} & {x>0} \\ {1} & {\left|\mathbf{r}-\mathbf{r}_{i j}\right|>r_{c}}\end{array}\right.
\end {equation}
where $ \mathbf{r}_{i j} $ is the position vector of the rod’s center. $ \left|\mathbf{r}-\mathbf{r}_{i j}\right|<r_{c} $ denotes the rod domain, and $ \left|\mathbf{r}-\mathbf{r}_{i j}\right|>r_{c} $ denotes the air domain. Note that $ \varepsilon^{(0)}(\mathbf{r}) $ has imaginary parts and the system is non-Hermitian. 

The Bloch states for the interface states of PC with $ \varepsilon^{(0)}(\mathbf{r}) $ can be expressed as $ E_{z, \mathrm{\textbf{k}n}}^{(0)}(\mathbf{r})=u_{\mathbf{k} n}^{(0)}(\mathbf{r}) e^{i \mathbf{k} \cdot \mathbf{r}} $, where $ n $ denotes the band index and \(\textbf{k}\) is the wave vector in the first Brillouin zone. The periodic function $ u_{\mathbf{k}n}^{(0)}(\mathbf{r}) $ and the corresponding eigenfrequency $ \omega_{\mathrm{\textbf{k}n}}^{(0)} $ can be obtained from COMSOL. The Bloch states of non-Hermitian PC should satisfy a biorthonormal relationship as below 
\begin{equation}
\int {{{\rm{d}}^2}r\;} v_{{\bf{k}}m'}^{\left( 0 \right)*}\left( {\bf{r}} \right){\varepsilon ^{\left( 0 \right)}}\left( {\bf{r}} \right)u_{{\bf{k}}m}^{\left( 0 \right)}\left( {\bf{r}} \right) = {\delta _{m'm}},
\end{equation}
where  $v_{{\bf{k}}m'}^{\left( 0 \right)}\left( {\bf{r}} \right)$ is the normalized left eigenfield and can be obtained by the normalized right eigenfield at $ -\textbf{k} $, i.e.  $v_{{\bf{k}}m}^{\left( 0 \right)*}\left( {\bf{r}} \right) = u_{ - {\bf{k}}m}^{\left( 0 \right)}\left( r \right)$ [See Appendix \ref{sec:appB}].

 To construct the model Hamiltonian for a PC with a new permittivity $ \varepsilon(\mathbf{r}) $, which has the same functional form as Eq. \eqref{eq:c5-parameterepsilon}.  we express Bloch wave functions of the new PC as $  E_z(\mathbf{r})=E_{z, \mathbf{k}n}(\mathbf{r})=u_{\mathbf{k} n}(\mathbf{r}) e^{i \mathbf{k} \cdot \mathbf{r}} $ with $ u_{\mathbf{k} n}(\mathbf{r})=\sum_{m=1}^{+\infty} \alpha_{n, \mathbf{k} m} u_{\mathbf{k} m}^{(0)}(\mathbf{r}) $. Substituting this expansion into Eq. \eqref{eq:inter-Helmo}, we arrive at 
\begin{eqnarray}
\sum\limits_m^{} {{\alpha _{n,{\bf{k}}m}}} \left\{ \begin{array}{l}
{\nabla ^2}\left( {u_{{\bf{k}}m}^{\left( 0 \right)}\left( {\bf{r}} \right){e^{i{\bf{k}}r}}} \right)\\
 + \varepsilon \left( {\bf{r}} \right){\left( {\frac{{\omega _{{\bf{k}}n}^{}}}{{\rm{c}}}} \right)^2}u_{{\bf{k}}m}^{\left( 0 \right)}\left( {\bf{r}} \right){e^{i{\bf{k}}r}}
\end{array} \right\} = 0.
\label{eq:inter-Hmodelequation}
\end{eqnarray}
For the original PC with $\varepsilon _{}^{\left( 0 \right)}\left( {\bf{r}} \right)$, we know that 
\begin{equation}
	{\nabla ^2}\left( {u_{{\bf{k}}m}^{\left( 0 \right)}\left( {\bf{r}} \right){e^{i{\bf{k}}r}}} \right) + \varepsilon _{}^{\left( 0 \right)}\left( {\bf{r}} \right){\left( {\frac{{\omega _{{\bf{k}}m}^{\left( 0 \right)}}}{{\rm{c}}}} \right)^2}u_{{\bf{k}}m}^{\left( 0 \right)}\left( {\bf{r}} \right){e^{i{\bf{k}}r}} = 0.
\end{equation}
Then Eq. \eqref{eq:inter-Hmodelequation} becomes 
\begin{equation}\label{eq:c5-hamiltonianmodelequation}
	\sum\limits_m^{} {{\alpha _{n,{\bf{k}}m}}} \left[ \begin{array}{l}
\varepsilon \left( {\bf{r}} \right){\left( {\frac{{\omega _{{\bf{k}}n}^{}}}{{\rm{c}}}} \right)^2}\\
 - \varepsilon _{}^{\left( 0 \right)}\left( {\bf{r}} \right){\left( {\frac{{\omega _{{\bf{k}}m}^{\left( 0 \right)}}}{{\rm{c}}}} \right)^2}
\end{array} \right]u_{{\bf{k}}m}^{\left( 0 \right)}\left( {\bf{r}} \right){e^{i{\bf{k}}r}} = 0,
\end{equation}
where $\omega_{\mathrm{\textbf{k}n}}$ is eigenfrequency for PCs with $\varepsilon(\mathbf{r}) $ .
Multiplying Eq. \eqref{eq:c5-hamiltonianmodelequation} by  $v_{{\bf{k}}m'}^{\left( 0 \right)*}\left( {\bf{r}} \right)$ and integrating within a unit cell, we obtain 
 \begin{eqnarray}\label{eq:c5-hamiltonianmodel2}
\begin{array}{l}
\sum\limits_{m = 1}^{ + \infty } {{\alpha _{n,{\bf{k}}m}}} {\left( {\bar \varepsilon _{}^{\left( 0 \right)}\left( {\bf{k}} \right)} \right)_{m'm}}{(\omega _{{\bf{k}}m}^{\left( 0 \right)}/c)^2}\\
 = {({\omega _{{\bf{k}}n}}/c)^2}\sum\limits_{m = 1}^{ + \infty } {{\alpha _{n,{\bf{k}}m}}} {\left( {{{\bar \varepsilon }_{}}\left( {\bf{k}} \right)} \right)_{m'm}}
\end{array}
 \end{eqnarray}
in which 
\begin{subequations}\label{eq:c5-matrixepsilon}
	\begin{eqnarray}
		&{\left( {{{\bar \varepsilon }^{\left( 0 \right)}}\left( {\bf{k}} \right)} \right)_{m'm}} = \int {{{\rm{d}}^2}r\;} v_{{\bf{k}}m'}^{\left( 0 \right)*}\left( {\bf{r}} \right){\varepsilon ^{\left( 0 \right)}}\left( {\bf{r}} \right)u_{{\bf{k}}m}^{\left( 0 \right)}\left( {\bf{r}} \right), \\
		&{\left( {\bar \varepsilon \left( {\bf{k}} \right)} \right)_{m'm}}
		= \int {{{\rm{d}}^2}r\;} v_{{\bf{k}}m'}^{\left( 0 \right)*}\left( {\bf{r}} \right)\varepsilon \left( {\bf{r}} \right)u_{{\bf{k}}m}^{\left( 0 \right)}\left( {\bf{r}} \right).
	\end{eqnarray}
%\begin{aligned}
%\end{aligned}
\end{subequations}
Therefore, Eq. (\ref{eq:c5-hamiltonianmodel2}) can be rewritten as a generalized eigenvaule problem for a specific \textbf{k} as 
\begin{equation}\label{eq:c5-generalizedeigenvalue}
H_{2} \cdot p_{\mathbf{k}n}=\left(\omega_{\mathbf{k}n} / c\right)^{2} H_{1} \cdot p_{\mathbf{k}n}
\end{equation}
where $ p_{\mathrm{\textbf{k}n}}=\left(\cdots \alpha_{n, \mathbf{k} n}, \cdots\right)^{T} $ is the eigenstate. The matrices in Eq. (\ref{eq:c5-generalizedeigenvalue}) are   $ \left(H_{1}\right)_{m^{\prime} m}=(\overline{\varepsilon}(\mathbf{k}))_{m^{\prime} m}  $ and $ \left(H_{2}\right)_{m^{\prime} m}=\delta_{m^{\prime} m}\left(\omega_{\mathrm{\textbf{k}m}}^{(0)} / c\right)^{2} $. For convenience, we omit the subscript \(\textbf{k}\) for simplicity and rewrite Eq.(\ref{eq:c5-generalizedeigenvalue}) as
\begin{equation}\label{eq:c5-generalizedeigenvalue2}
H \cdot p_{n}=W_{n} p_{n},
\end{equation}
where $ H=H_{1}^{-1} \cdot H_{2} $, and $ W_{n}=\left(\omega_{n} / c\right)^{2} $. From Eqs. (\ref{eq:c5-matrixepsilon}) and \eqref{eq:c5-generalizedeigenvalue}, we see that the model Hamiltonian is a function of $ \varepsilon(\textbf{r}) $ for a fixed $ \textbf{k}=k_y \hat y $. Therefore, using the eigenfunctions of the interface states of a system with permittivity $ \varepsilon^{(0)}(\mathbf{r}) $ at a specific $ k_y $ as bases, we build a Hamiltonian model to calculate a PC with new permittivity $ \varepsilon(\mathbf{r}) $. This approach works well for any value of $ k_y $.

\begin{figure}
	\centering
	\includegraphics[width=0.7\linewidth]{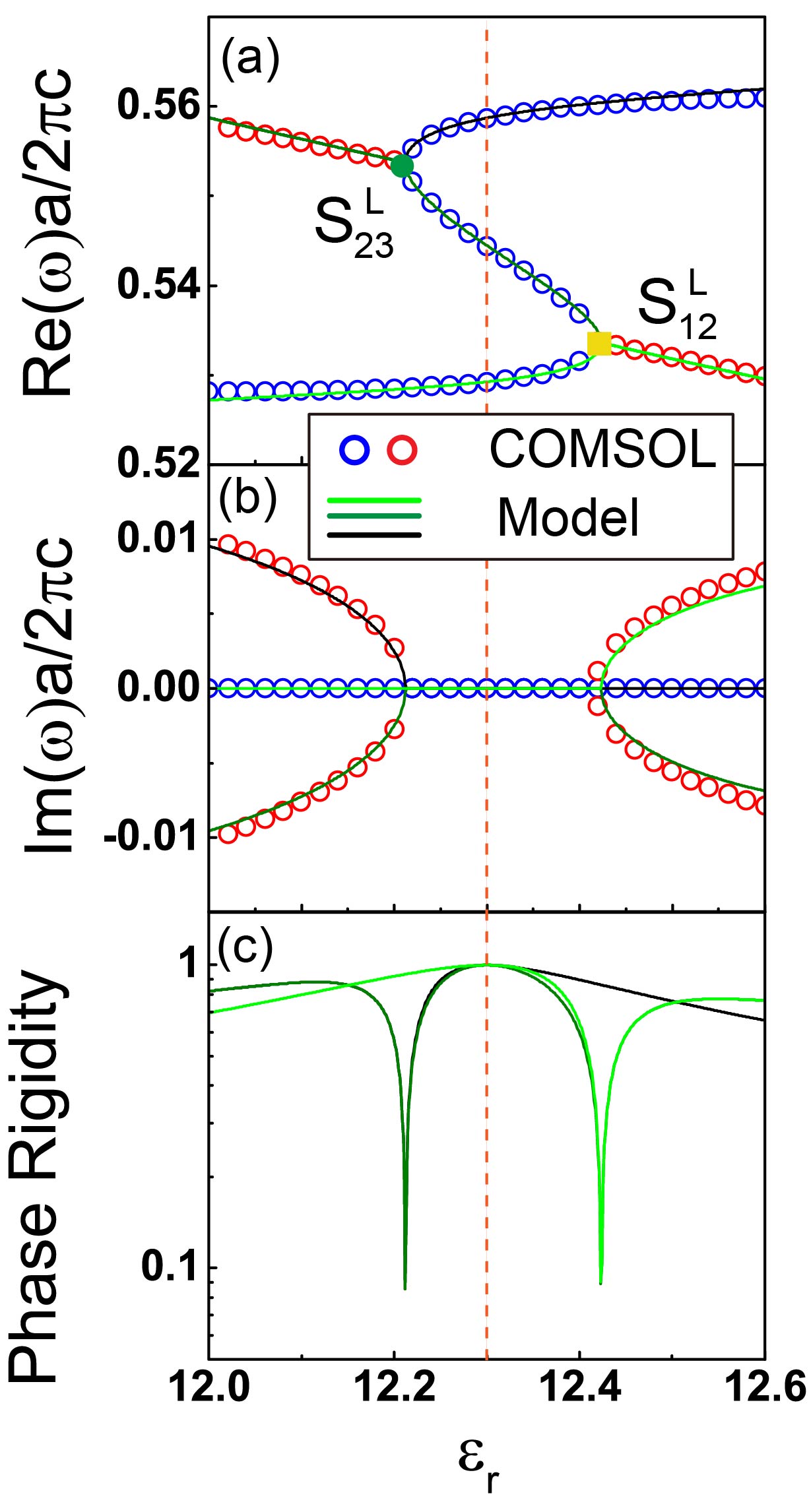}
	\caption{(Color online) (a) Real parts and (b) imaginary parts of eigenfrequencies of interface states (with a specific $ k_{y} a / 2 \pi=0.05 $) as functions of $ \varepsilon_r $ are plotted. The original parameters of the cylinders we used to build the model are $ \varepsilon_{r}^{(0)}=12.3 \text { and } \gamma^{(0)}=1.2 $, which are denoted by an orange dashed line. The 1$^ \text{st} $ , 2$^ \text{nd} $ and 3$^ \text{rd} $ bands calculated by the Hamiltonian model are plotted by green, olive and black solid lines, respectively. The phase rigidities $ |r_m| $ of the interface states as a function of $ \varepsilon_r $ are plotted in (c). The open circles in (a) and (b) are calculated by COMSOL with PML boundary conditions. The number of column layers of the semi-infinite PCs is truncated to $ N=15 $. Blue circles represent states in exact $\mathcal{PT}$-symmetry phase, while red circles represent states in broken $\mathcal{PT}$-symmetry phase.}
	\label{fig:model_epsilonr}
\end{figure}

Using the model Hamiltonian, we can analyze the dispersion of interface states as functions of the material parameters $ \gamma $ and $ \varepsilon_r $. We first consider the transition of the interface band dispersion depicted in Figs. \ref{fig:EMTandBandinversion}(b) and \ref{fig:EMTandBandinversion}(d), in which $ \varepsilon_r $ changes from 12 to 12.6. We calculate the interface states with parameters $ \varepsilon_{r}^{(0)}=12.3 $ and $ \quad \gamma^{(0)}=1.2 $ using COMSOL, and focus on the three interface states with real eigenvalues at a specific $ k_ya/2\pi=0.05 $ (orange dashed lines in Fig. \ref{fig:model_epsilonr}(a)). We use the eigenfunctions $ u_{\mathbf{k} n}^{(0)}(\mathbf{r}) $ of these three interface states as bases to build the Hamiltonian model shown in Eq. \eqref{eq:c5-generalizedeigenvalue2}. The Hamiltonian $ H $ is a $ 3 \times 3 $ matrix function of $ \varepsilon_r $  and $ \gamma $. We fixed the non-Hermiticity at $ \gamma=\gamma^{(0)}=1.2 $, and calculate the eigenvalues of the Hamiltonian $ H $ as a function of $ \varepsilon_r $. The real and imaginary parts of the eigenfrequencies are shown respectively in Figs. \ref{fig:model_epsilonr}(a) and \ref{fig:model_epsilonr}(b) by solid lines. Ordered from lower frequency to higher frequency, the 1$^ \text{st} $ (lowest), 2$^ \text{nd} $ (middle), and 3$^ \text{rd} $ (highest) bands are denoted by green, olive and black lines respectively. We also plot the numerical results calculated directly using COMSOL with PML boundary conditions by open circles. The blue circles denote the states in exact $\mathcal{PT}$ symmetry phase (the eigenvalues are purely real) and the red circles denote the states in broken $\mathcal{PT}$-symmetry phase (the eigenvalues are complex conjugate pairs). The solid lines show excellent agreement with the open circles, indicating the validity of our non-Hermitian $ 3 \times 3 $ model Hamiltonian.

The 2$^ \text{nd} $ and 3$^ \text{rd} $ bands merge together at $ \varepsilon_{r}=12.212 $ and form an EP marked by $ S_{23}^L $ (green dot) in Fig. \ref{fig:model_epsilonr}(a). The 1$^ \text{st} $ band and the 2$^ \text{nd} $ band merge at $ \varepsilon_{r}=12.423 $, and form an EP marked by $ S_{12}^L $ (yellow square). The subscript index $ mn $ of $ S_{23}^L $ and $ S_{12}^L $  denotes the index of bands forming the EPs. Using the $ 3 \times 3 $ Hamiltonian, we can analyze the EPs in the $ \varepsilon_r $ parameter space in detail. The eigenstates of the non-Hermitian Hamiltonian matrix become defective at the EP, characterized by a vanishing phase rigidity which is defined as $ {r_m} =  \left\langle {v_{{\bf{k}}m}^{}} \right|\left. {u_{{\bf{k}}m}^{}} \right\rangle /\left\langle {u_{{\bf{k}}m}^{}} \right|\left. {u_{{\bf{k}}m}^{}} \right\rangle $, where $\left| {u_{{\bf{k}}m}^{}} \right\rangle $ and $\left| {v_{{\bf{k}}m}^{}} \right\rangle $ are the right and left eigenstates \cite{Bulgakov2006}. We plot the phase rigidity as a function of $\varepsilon_r $ in Fig. \ref{fig:model_epsilonr}(c). The vanishing phase rigidities, $ r_m \to 0 $, also confirms the existence of the EPs. 

Using the Hamiltonian model, we described the EPs in the parameter space of $ \varepsilon_r $ for a fixed $ k_{y} a / 2 \pi=0.05$ in Fig. \ref{fig:model_epsilonr}. Now we will show the turning points denoted by a green dot and a yellow square in Fig. \ref{fig:EMTandBandinversion}(b) and \ref{fig:EMTandBandinversion}(d) are EPs in the parameter space of $ k_y $. In Fig. \ref{fig:EMTandBandinversion}(b), where $ \varepsilon_r=12 $, the lowest band emerges from the Brillouin zone center with a negative group velocity and the middle and highest bands form an EP $ S_{23}^L $. In Fig. \ref{fig:model_epsilonr}, the EP $ S_{23}^L $ appears at $ \varepsilon_r=12.212 $ meaning that we can find an EP $ S_{23}^L $ at $ k_{y} a / 2 \pi=0.05$ for PC with  $ \varepsilon_r=12.212 $. Similarly, in Fig. \ref{fig:model_epsilonr} an EP $ S_{12}^L $ appears at $ \varepsilon_r=12.423 $ meaning that we can find EP $ S_{12}^L $ at $ k_{y} a / 2 \pi=0.05$ for PC with  $ \varepsilon_r=12.423 $. In Fig. \ref{fig:EMTandBandinversion}(d), where $ \varepsilon_r=12.6 $, the lower two bands form an EP $ S_{12}^L $, and the highest band emerges from the Brillouin zone center with a positive group velocity.
\begin{figure}
	\centering
	\includegraphics[width=\linewidth]{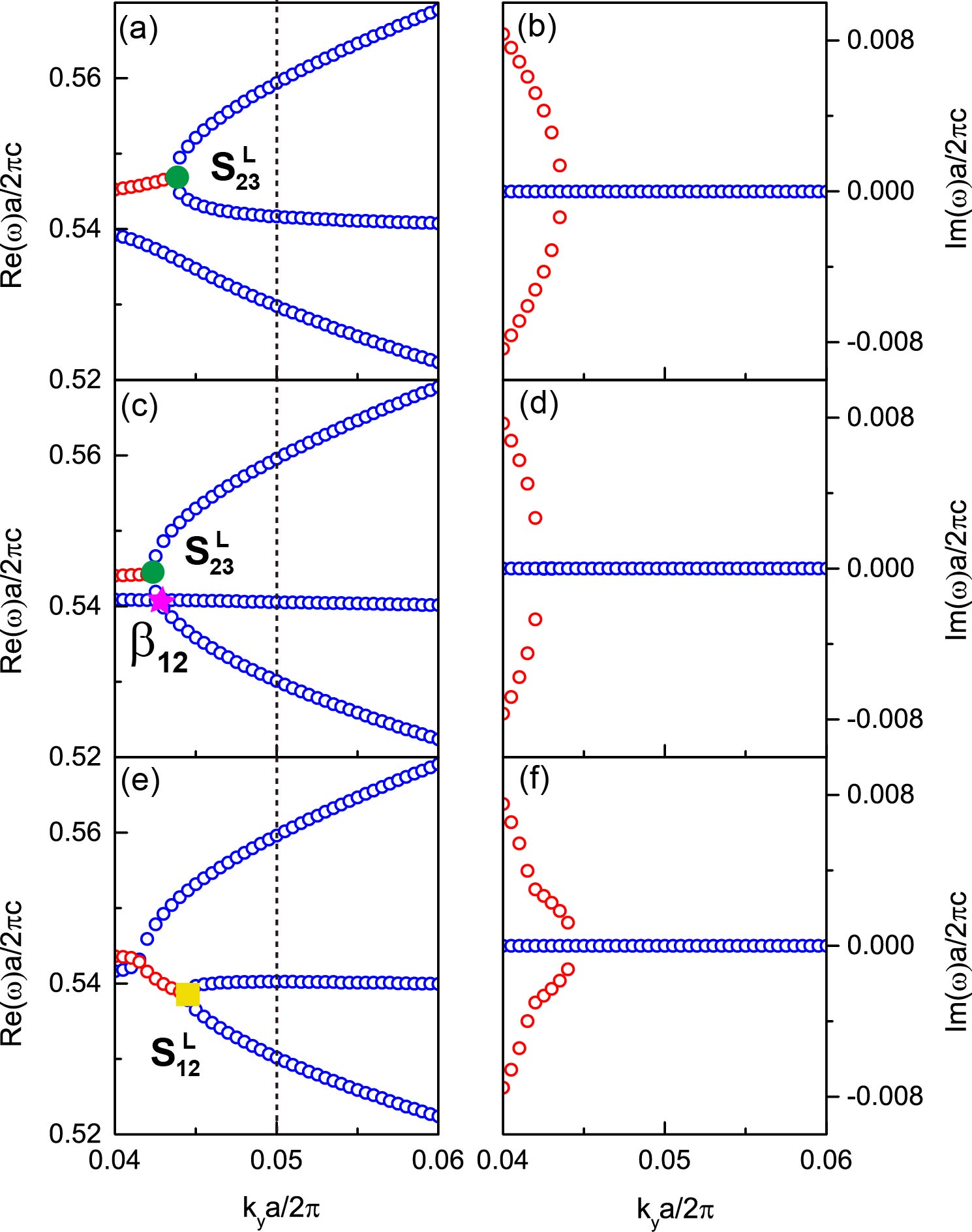}
	\caption{(Color online) (a),(c),(e) Real and (b),(d),(f) imaginary parts of eigenfrequencies of interface states as functions of $ k_y $ are calculated by COMSOL with PML boundary conditions. The number of column layers of the semi-infinite PCs is truncated to $ N=15 $. The blue (red) circles represent states in exact (broken) $\mathcal{PT}$-symmetry phase. The relative permittivities of the cylinders are (a-b) $ \varepsilon_{c}=12.34 \pm 1.2 i $; (c-d) $ \varepsilon_{c}=12.3553 \pm 1.2 i $; and (e-f) $  \varepsilon_{c}=12.36 \pm 1.2 i $. The green dots, solid yellow square and pink star labeled by $ S_{23}^L $, $ S_{12}^L $ and $ \beta_{12} $ are EPs. The vertical black dashed line labels ${k_y}a/2\pi  = 0.05$. }
	\label{fig:bandinversionbeta}
\end{figure}

\begin{figure}
	\centering
	\includegraphics[width=\linewidth]{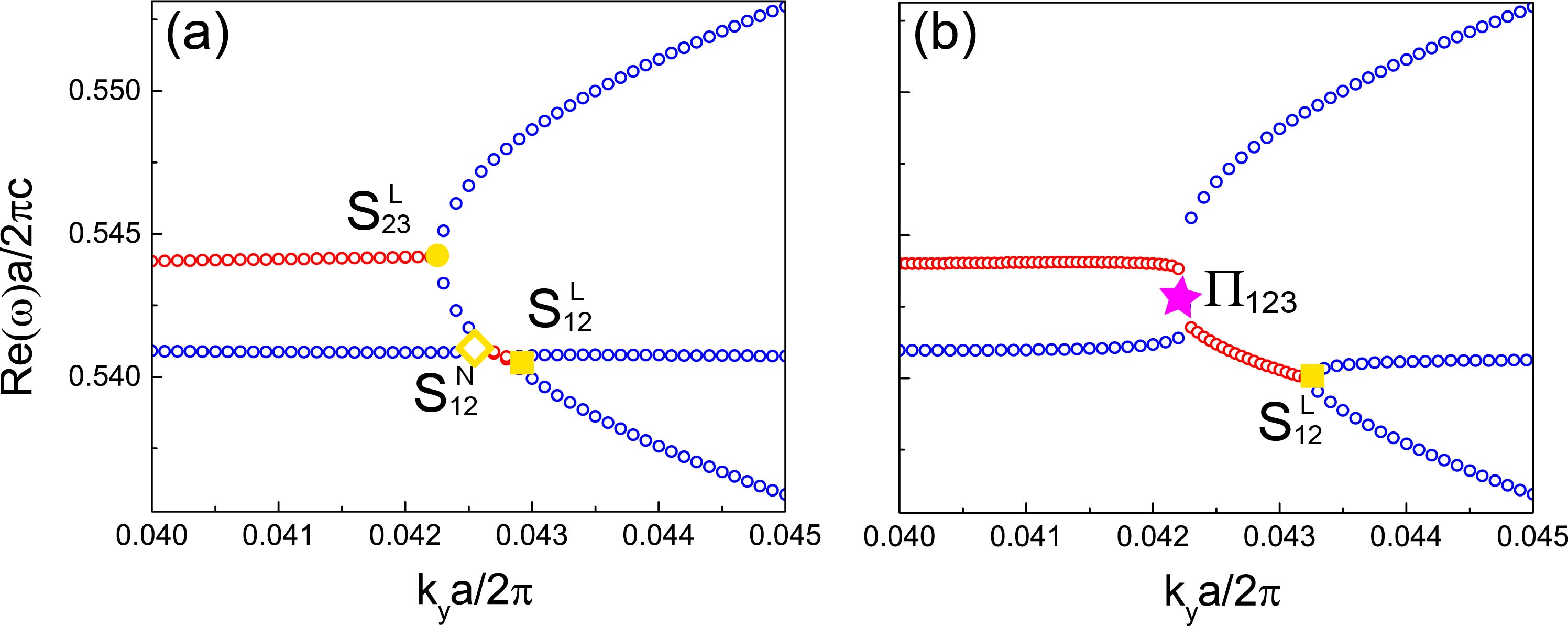}
	\caption{ (Color online) Real parts of eigenfrequencies of states as functions of $ k_y $ are calculated by COMSOL with PML boundary conditions for PC with (a) $ \varepsilon_{c}=12.3554 \pm 1.2 i $, (b) $ \varepsilon_{c}=12.3559 \pm 1.2 i $. The number of column layers of the semi-infinite PCs is truncated to $ N=15 $. The blue (red) circles represent states in exact (broken) $\mathcal{PT}$-symmetry phase. The yellow dot, yellow solid squares, yellow open diamond and pink star labeled by $ S_{23}^L $, $S_{12}^L  $, $ S_{12}^N $ and  $ \Pi_{123} $ are EPs. }
	\label{fig:bandinversionbetaPi}
\end{figure}

To better describe the movement of EPs in $ \varepsilon_r $ and $ k_y $ parameter spaces, in Fig. \ref{fig:bandinversionbeta}, we plot the interface bands of PC with (a-b) $ \varepsilon_{c}=12.34 \pm 1.2 i $, (c-d) $ \varepsilon_{c}=12.3553 \pm 1.2 i $, and (e-f) $  \varepsilon_{c}=12.36 \pm 1.2 i $, respectively. From Figs. \ref{fig:bandinversionbeta}(a) and \ref{fig:bandinversionbeta}(c), we see that as $ \varepsilon_r $ increases, the middle band get closer to the lowest band and they merge together at $ \varepsilon_r=12.3553 $. This touching point denoted by $ \beta_{12} $ is an anisotropic EP, which we will discuss in next section. As $ \varepsilon_r $ increases further, as shown in Fig. \ref{fig:bandinversionbeta}(e), the EP $ S_{23}^L $ disappears and an EP $ S_{12}^L $ formed by the lower two bands comes out from the touching point $ \beta_{12} $.
Figure \ref{fig:bandinversionbetaPi} describes this process. We can see that the touching point $ \beta_{12} $ first split into two EPs $ S_{12}^N $ and $ S_{12}^L $, and then the EPs $ S_{12}^N $ and $ S_{23}^L $ coalesce into an order-3 EP $ \Pi_{123} $, where three bands coalesce \cite{Hodaei2017}.
In summary, Figs. \ref{fig:bandinversionbeta} and \ref{fig:bandinversionbetaPi} describe the transition from $ S_{23}^L $ to $ S_{12}^L $, and show that as $ \varepsilon_r $ changes, the evolution of the interface bands is related to the development of EPs.

\section{\label{sec:6}Coalescence of exceptional points}
\begin{figure}
	\centering
	\includegraphics[width=\linewidth]{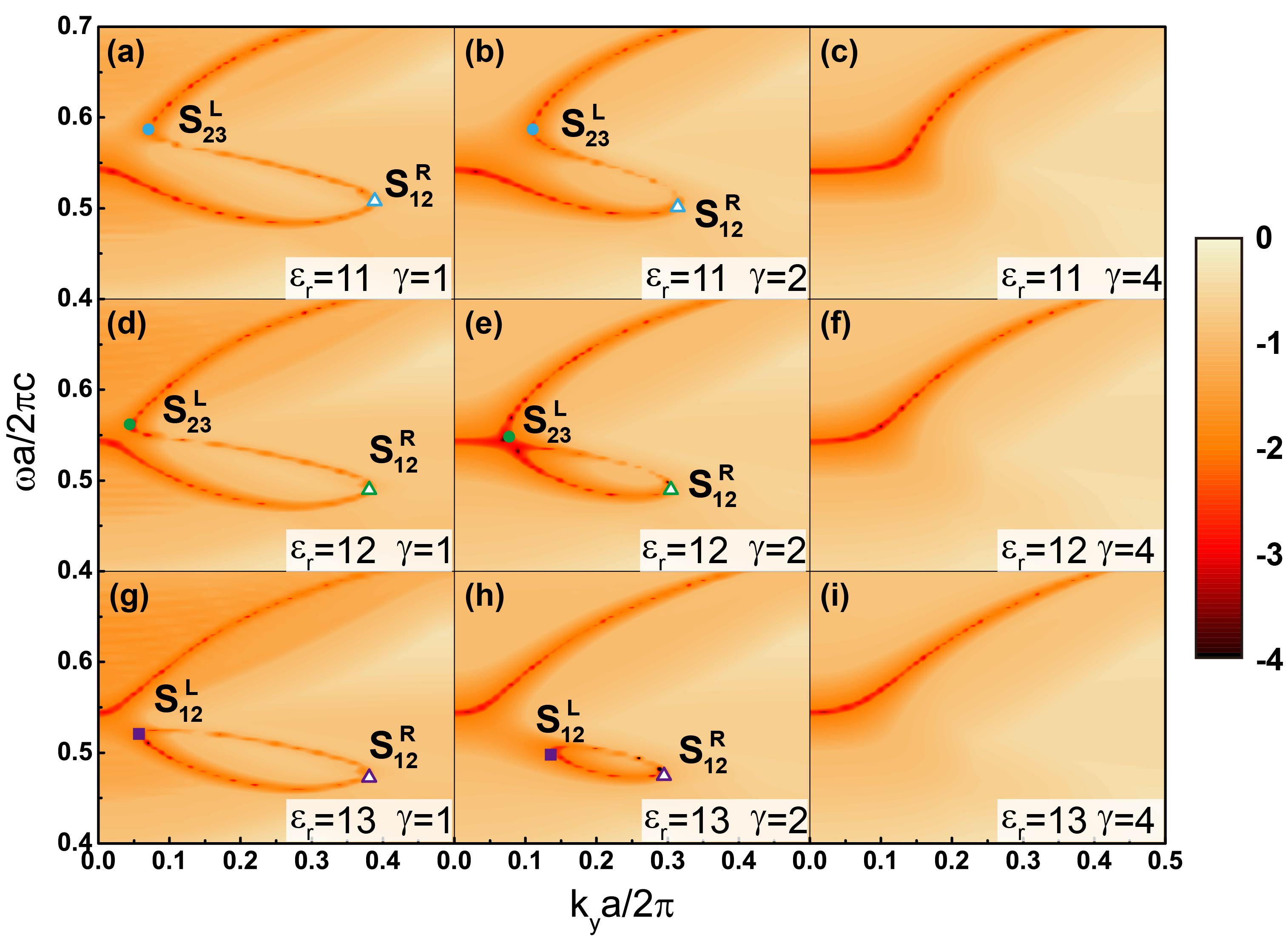}
	\caption{(Color online) Colormap plot of \({\rm{Min}}\left\{ {\left| {{\mathop{\rm eig}\nolimits} \left[ {{M}\left( {\omega ,{k_y}} \right)} \right]} \right|} \right\}\) in log-10 scale as functions of real frequency $ \omega $ and real Bloch wave number $ k_y $ for non-Hermitian PCs with different parameters $ \varepsilon_r $ and $ \gamma $. The dark orange lines denote the $\mathcal{PT}$-symmetric interface states. The relative permittivities of cylinders are labeled in the figure panels and the number of column layers of the semi-infinite PCs is truncated to $ N=15 $. Different symbols labeled by $ S_{23}^L $, $ S_{12}^L $ and $ S_{12}^R $ are EPs.}
	\label{fig:interfacebandpattern}
\end{figure}

\begin{figure}
	\centering
	\includegraphics[width=\linewidth]{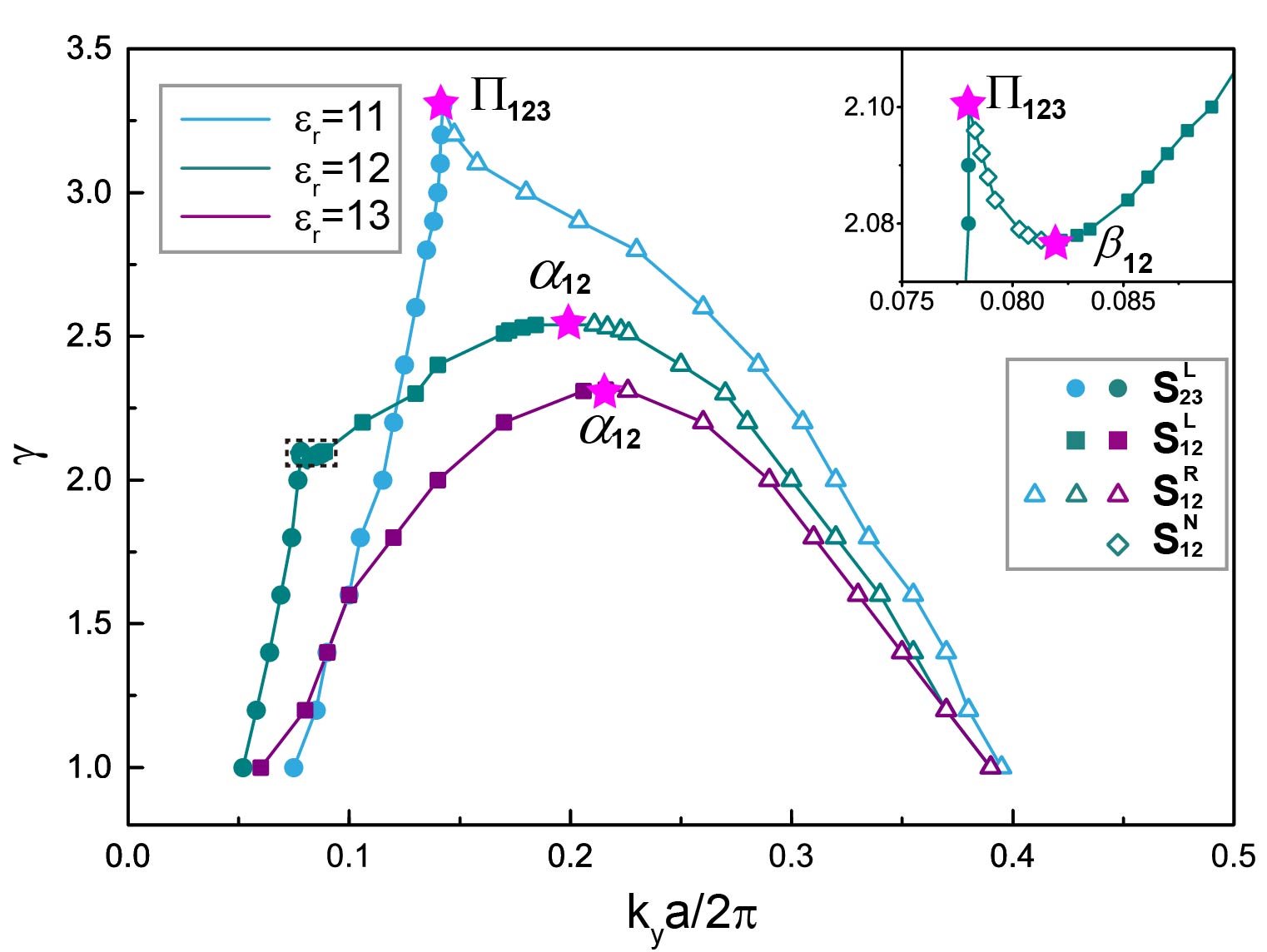}
	\caption{(Color online) The trajectories of EPs in the parameter space $ (k_y, \gamma) $ for PCs with different real parts of relative permittivity $ \varepsilon_r $. Blue, green and purple lines represent PCs with  $ \varepsilon_r=11 $, $ \varepsilon_r=12 $, and $ \varepsilon_r=13 $, respectively. Different symbols (dots, solid squares, open triangles, and open diamonds) label the order-2 EPs ($ S_{23}^L $, $ S_{12}^L $, $ S_{12}^R $ and $ S_{12}^N $). The pink stars denote the coalescence of EPs. EPs marked by $ \Pi_{123} $ are order-3 EPs. $ \alpha_{12} $ and $ \beta_{12} $ are anisotropic EPs. The results are calculated by COMSOL with PML boundary conditions and the number of column layers of the semi-infinite PCs is truncated to $ N=15 $. The upper right inset is the enlargement of the black dashed rectangle.}
	\label{fig:EPsinkgamma}
\end{figure}

In the above section, we build a $ 3 \times 3 $ non-Hermitian Hamiltonian model to analyze the dispersion of interface states of $\mathcal{PT}$-symmetric photonic crystals as functions of the material parameters $ \varepsilon_c $, and describe the movement of EPs in parameter space $ \varepsilon_r $ and $ k_y $. But we did not take the change of the non-Hermiticity $ \gamma $ into account. In this section, we expand the parameter space and study systematically the evolution of interface states as $ \varepsilon_r $  and $ \gamma $ changes. Interface bands calculated using the MS method with open boundary conditions are plotted in Fig. \ref{fig:interfacebandpattern} for different parameters  $ \varepsilon_c=\varepsilon_r\pm i \gamma $, with $ \varepsilon_r $ varying from 11 to 13, and $ \gamma  $ varying from 1 to 4. The interface bands show ziz-zag or closed-loop dispersion that cannot be seen in Hermitian systems. The turning points of the bands are EPs which are labeled by the symbols $ S_{23}^L $, $ S_{12}^L $ and $ S_{12}^R $ in the figure panels. In this following, we will show that the evolution of the interface bands is closely associated with the coalescence of the EPs as we change the parameters $ \varepsilon_r $ and $ \gamma $. Such coalescence behavior is further highlighted by tracing EPs in the parameter space $ (k_y, \gamma) $, as shown in Fig. \ref{fig:EPsinkgamma}. The last column of panels in Fig. \ref{fig:interfacebandpattern} shows that when gain/loss becomes large, one isolated band of interface states will always persists. 

\subsection{Formation of the order-3 EPs in $ \varepsilon_r=11 $}

In the first row of panels in Fig. \ref{fig:interfacebandpattern}, the real part of the relative permittivity of the cylinders is fixed at $ \varepsilon_r=11 $. As shown in Fig. \ref{fig:interfacebandpattern}(a), where $ \gamma=1 $, the band of interface states starts from the Brillouin zone center and exhibits a ziz-zag dispersion, turning around twice at EPs denoted by $ S_{23}^L $ and $ S_{12}^R $. As we increase the non-Hermitian strength $ \gamma $, the EPs $ S_{23}^L $ and $ S_{12}^R $ will get closer to each other and eventually disappear as shown in Fig. \ref{fig:interfacebandpattern}(b-c). This process is also illustrated in Fig. \ref{fig:EPsinkgamma} by the blue line, which traces the movement of EPs in the parameter space $ (k_y, \gamma) $.  We can see that as $ \gamma $ increases, the EPs $ S_{23}^L $ marked by dots at small $ k_y $ and EPs $ S_{12}^R $ marked by open triangles at large $ k_y $ get closer and eventually merge into an order-3 EP labeled as $ \Pi_{123} $  (marked by a pink star) \cite{Demange2012}. The EPs $ S_{23}^L $ and $ S_{12}^R $ are order-2 EPs, where two bands coalesce. At the order-3 EP $ \Pi_{123} $, three bands coalesce.

\begin{figure}
	\centering
	\includegraphics[width=\linewidth]{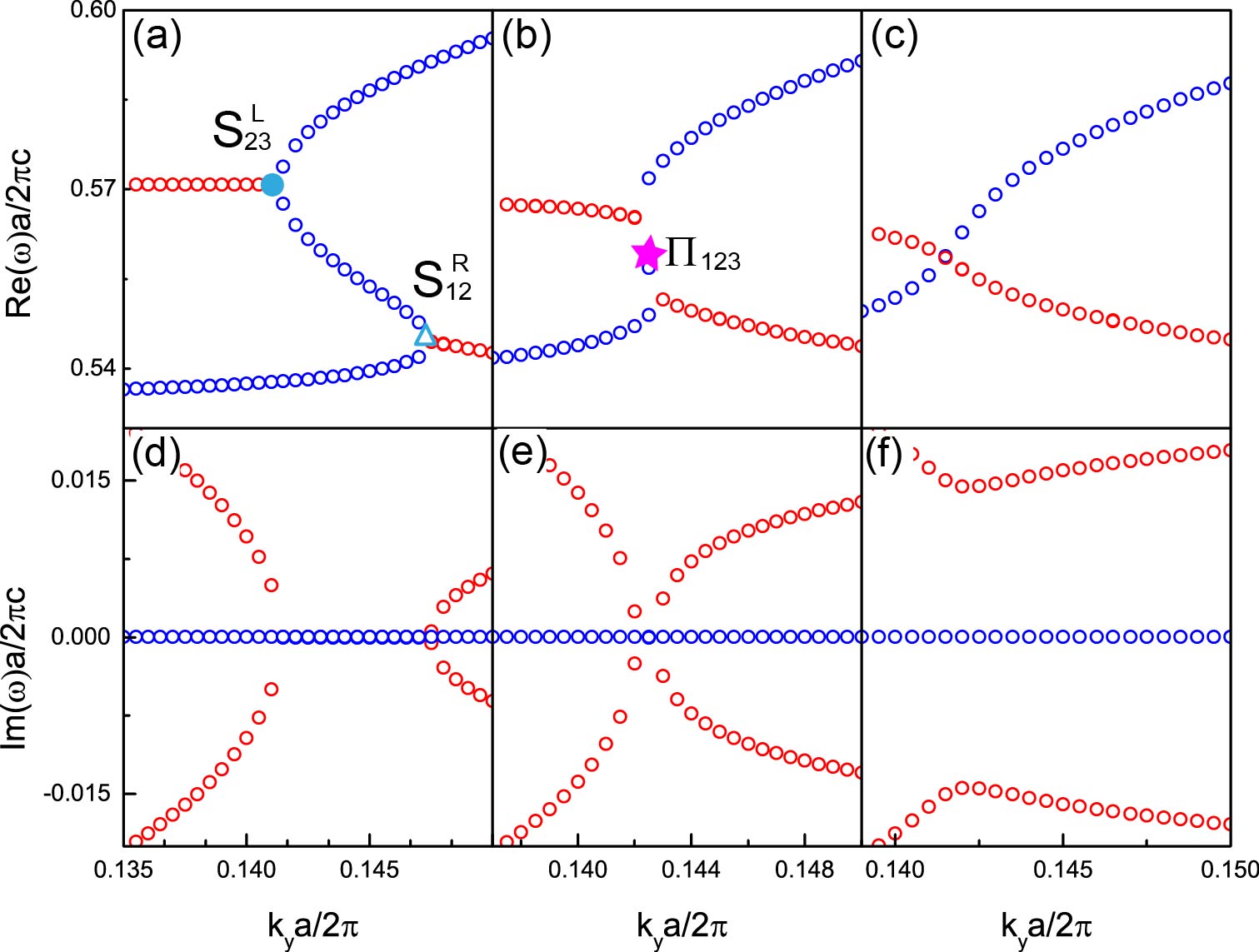}
	\caption{(Color online) (a-c) Real parts and (d-f) imaginary parts of eigenfrequencies of interface states as functions of $ k_y $ are calculated by COMSOL with PML boundary condition. The number of column layers of the semi-infinite PCs is truncated to $ N=15 $. The blue (red) circles represent interface states in exact (broken) $\mathcal{PT}$-symmetry phase. The relative permittivities of the cylinders are (a), (d) $ \varepsilon_{c}=11 \pm 3.2 i $ ; (b), (e) $ \varepsilon_{c}=11 \pm 3.3 i $ ; and (c), (f) $ \varepsilon_{c}=11 \pm 3.4 i $. The symbols labeled by $ S_{23}^L $, $ S_{12}^R $ and $ \Pi_{123} $ are EPs. }
	\label{fig:thirdorderEPs}
\end{figure}

To further illustrate the formation of the order-3 EP $ \Pi_{123} $,  we plot in Fig. \ref{fig:thirdorderEPs} the interface bands calculated using COMSOL with PML boundary conditions applied on the $ x $-direction. As shown in Figs. \ref{fig:thirdorderEPs}(a) and \ref{fig:thirdorderEPs}(d), when $ \varepsilon_c=11 \pm 3.2i $, there are two typical order-2 EPs $ S_{23}^L $ (blue dot) and $ S_{12}^R $ (open triangle) formed by two bands. The dispersion relations for the real and imaginary branches are of the square-root form near the EPs. As we increase the non-Hermiticity to $ \gamma=3.3 $, as shown in Figs. \ref{fig:thirdorderEPs}(b) and \ref{fig:thirdorderEPs}(e), the EPs $ S_{23}^L $ and $ S_{12}^R $ merge into an order-3 EP $ \Pi_{123} $, where three bands coalesce. The EP $ \Pi_{123} $ disappears upon a further increase in $\gamma$. As shown in Figs. \ref{fig:thirdorderEPs}(c) and \ref{fig:thirdorderEPs}(f), when $ \gamma=3.4 $, there is only one interface state with real eigenvalues (blue open circles). Other interface states (red open circles) will acquire larger and larger imaginary parts and merge into the continuum of propagating waves that are not localized at the gain-loss interface.

\subsection{Formation of the anisotropic EPs in  $ \varepsilon=13 $}

\begin{figure}
	\centering
	\includegraphics[width=\linewidth]{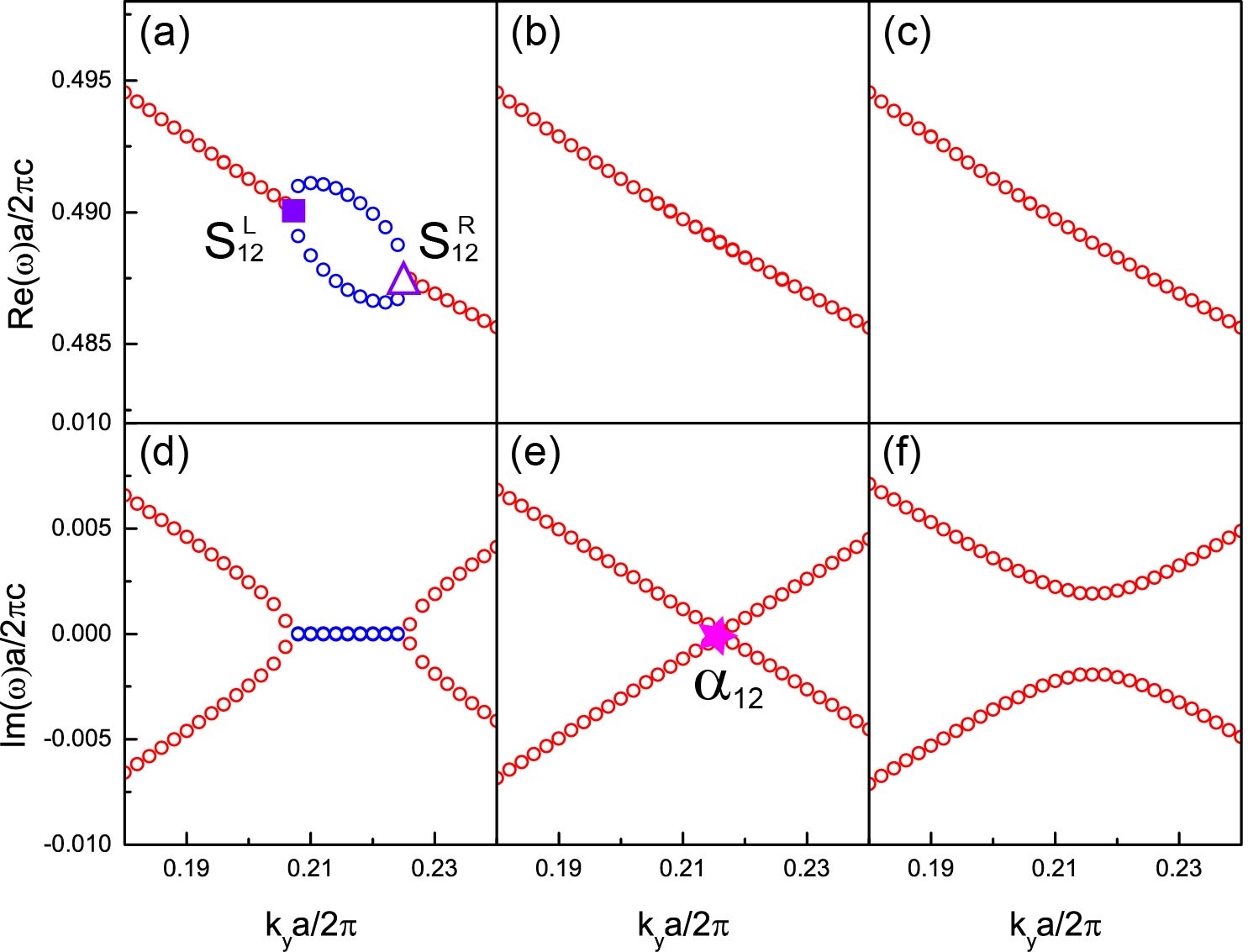}
	\caption{(Color online) (a-c) Real parts and (d-f) imaginary parts of eigenfrequencies of interface states as functions of $ k_y $ are calculated by COMSOL with PML boundary condition. The number of column layers of the semi-infinite PCs is truncated to $ N=15 $. The blue (red) circles represent interface states in exact (broken) $\mathcal{PT}$-symmetry phase. The relative permittivities of the cylinders are (a), (d) $ \varepsilon_{c}=13 \pm 2.31 i $ ; (b), (e) $ \varepsilon_{c}=13 \pm 2.3148 i $ ; and (c), (f) $ \varepsilon_{c}=13 \pm 2.32 i $. The symbols labeled by $ S_{12}^L $, $ S_{12}^R $ and $ \alpha_{12} $ are EPs. }
	\label{fig:anisotropicEPs}
\end{figure}

In the third row of panels in Fig. \ref{fig:interfacebandpattern}, the real part of the relative permittivity of the cylinders is increased to $ \varepsilon=13 $. In Fig. \ref{fig:interfacebandpattern} (g), we see that the lower two bands form a closed-loop, with the minimum and maximum \(k_y\) pinned by the EPs $ S_{12}^L $ and $ S_{12}^R $ respectively. From Figs. \ref{fig:interfacebandpattern}(g-i), we observe that when $ \gamma $ increases, the two EPs $ S_{12}^L $ and $ S_{12}^R $ of the loop tend to approach each other and finally disappear. The disappearance of the loop is described in Fig. \ref{fig:anisotropicEPs}. EPs $ S_{12}^L $ and $ S_{12}^R $ coalesce at a specific $ k_y $ point and form a new EP denoted by $ \alpha_{12} $. We also plot the trace of EPs in the parameter space $ (k_y, \gamma) $ in Fig. \ref{fig:EPsinkgamma} by purple line. As $ \gamma $ increases, the EPs $ S_{12}^L $ labeled by solid squares and EPs $ S_{12}^R $ labeled by open triangles merge into a new EP labeled by $ \alpha_{12} $.

\begin{figure}
	\centering
	\includegraphics[width=0.8\linewidth]{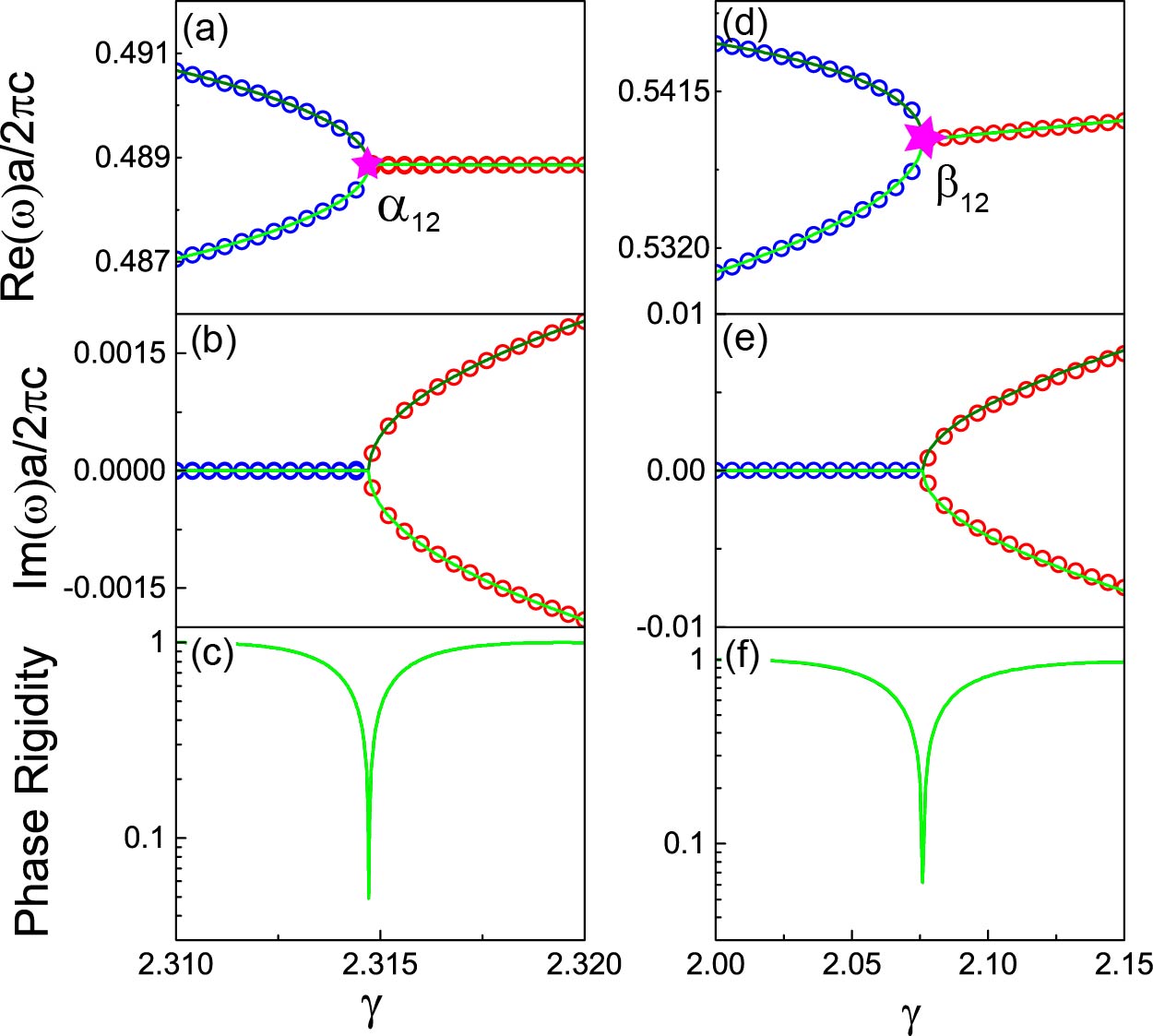}
	\caption{(Color online) (a), (d) Real parts, (b), (e) imaginary parts of eigenfrequencies and (c), (f) phase rigidities $ |r_m| $ of interface states as functions of $ \gamma $. The original parameters of the cylinders we used to build the Hamiltonian model for (a-c) is $ \varepsilon_{r}^{(0)}=13, \gamma^{(0)}=2.31$, $ k_{y} a / 2 \pi=0.216 $; and for (d-f) is $ \varepsilon_{r}^{(0)}=12$, $ \gamma^{(0)}=2 $, $ k_{y} a / 2 \pi=0.0818 $. Solid lines with different colors representing different interface bands are calculated by the Hamiltonian model. The circles are calculated by COMSOL with PML boundary conditions. The pink stars denote the EPs labeled as $ \alpha_{12} $ and $ \beta_{12} $. }
	\label{fig:anisotropicbetaalpha}
\end{figure}

It is interesting to note that EP $ \alpha_{12} $ is anisotropic \cite{Ding2018}. When an anisotropic EP is approached from different directions in the parameter space, it shows different singular behaviors. As shown in Fig. \ref{fig:anisotropicEPs}(e), the imaginary parts of the eigenfrequencies are linear functions of $ k_y $ near the EP $ \alpha_{12} $. In Figs. \ref{fig:anisotropicbetaalpha} (a-c), we plot the eigenfrequencies and phase rigidities of the interface states as a function of $ \gamma $ in the vicinity of the EP $ \alpha_{12} $. When the EP $ \alpha_{12} $ is approached along the $ \gamma $ direction, the dispersion shows a square-root behavior, in contrast to the linear behavior along $ k_y $ direction shown in Fig. \ref{fig:anisotropicEPs}(e). 

\subsection{Order-3 EPs and anisotropic EPs in $ \varepsilon_r=12 $ }

 In Fig. \ref{fig:interfacebandpattern}, the first row of panels shows the formation of an order-3 EP $ \Pi_{123} $, and the third row shows the formation of an anisotropic EP $ \alpha_{12} $. We now turn to the second row of panels in Fig. \ref{fig:interfacebandpattern}, where $ \varepsilon_r=12 $ and find that in addition to the formation of an order-3 EP, there is also formation of anisotropic EPs. We plot the trace of EPs in the parameter space $ (k_y, \gamma) $ in Fig. \ref{fig:EPsinkgamma} by green line. When $ \gamma $ is small, there are two typical EPs, one is the EP $ S_{23}^L $ labeled by solid dots at small $ k_y $ and the other is EP $ S_{12}^R $ labeled by open triangles at large $ k_y $. The corresponding interface state bands with two EPs and a zigzag dispersion is shown in Fig. \ref{fig:interfacebandpattern}(d). As shown in the right inset of Fig. \ref{fig:EPsinkgamma}, when we keep on increasing $ \gamma $, a new EP denoted by $ \beta_{12} $ appears, which we will discuss later, and then it splits into two EPs labeled as $ S_{12}^N $ (marked by open diamond) and $ S_{12}^L $ (marked by solid square). When $ \gamma $ increases further, the EPs $ S_{12}^N $ and $ S_{23}^L $ will coalesce into an order-3 EP labeled as $ \Pi_{123} $, which is similar to the formation of $ \Pi_{123} $ in the blue lines. This EP $ \Pi_{123} $ disappears as $ \gamma $ increases further, and there are only two EPs $ S_{12}^L $ and $ S_{12}^R $ left. These two EPs will coalesce into an anisotropic EP $ \alpha_{12} $ as $ \gamma $ increases, which is the same as the process we have illustrated in $ \varepsilon_r=13 $ denoted by the purple line.

  \begin{figure}
 	\centering
 	\includegraphics[width=\linewidth]{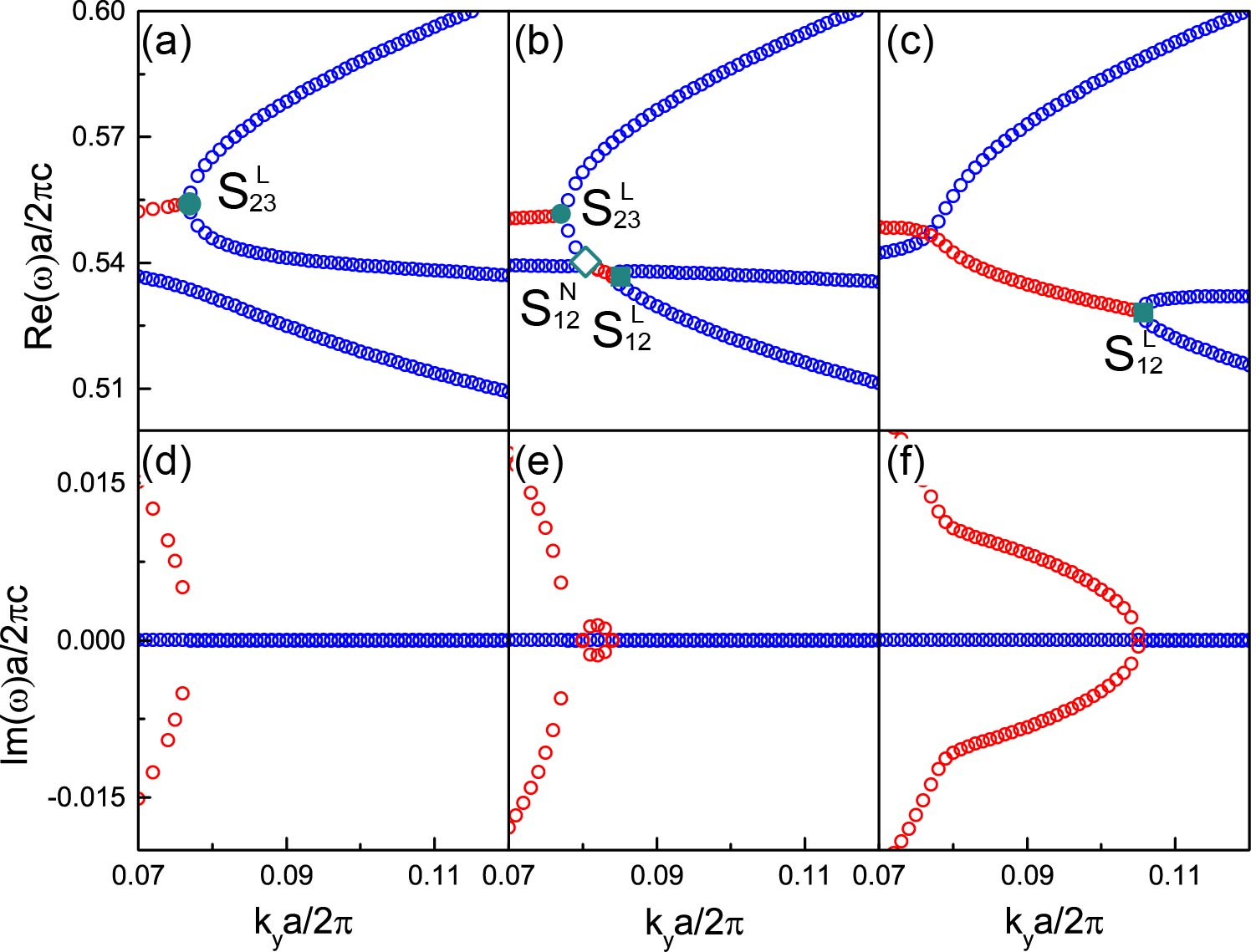}
 	\caption{(Color online) (a-c) Real parts and (d-f) imaginary parts of eigenfrequencies of interface states as functions of $ k_y $ are calculated by COMSOL with PML boundary condition. The number of column layers of the semi-infinite PCs is truncated to $ N=15 $. The blue (red) circles represent interface states in exact (broken) $\mathcal{PT}$-symmetry phase. The relative permittivities of the cylinders are (a), (d) $ \varepsilon_{c}=12 \pm 2 i $ ; (b), (e) $ \varepsilon_{c}=12 \pm 2.08 i $ ; and (c), (f) $ \varepsilon_{c}=12 \pm 2.2 i $. The symbols labeled by $ S_{23}^L $, $ S_{12}^L $ and $  S_{12}^N  $ are EPs. }
 	\label{fig:thirdorderandanisotripicEPs}
 \end{figure}

Comparing these three lines in Fig. \ref{fig:EPsinkgamma}, we find that the green line ($ \varepsilon_r=12 $) is more than just a composite of the blue line ($\varepsilon_r=11$) and the purple line ($\varepsilon_r=13$). In the green line, we find the formation of an order-3 EP $ \Pi_{123} $ which also appears in the blue line, and the formation of an anisotropic EP $ \alpha_{12} $ which also appears in the purple line. But the splitting of an EP $ \beta_{12}$ illustrated in the inset of Fig. \ref{fig:EPsinkgamma} is unique to the green line.

\begin{figure}
	\centering
	\includegraphics[width=0.7\linewidth]{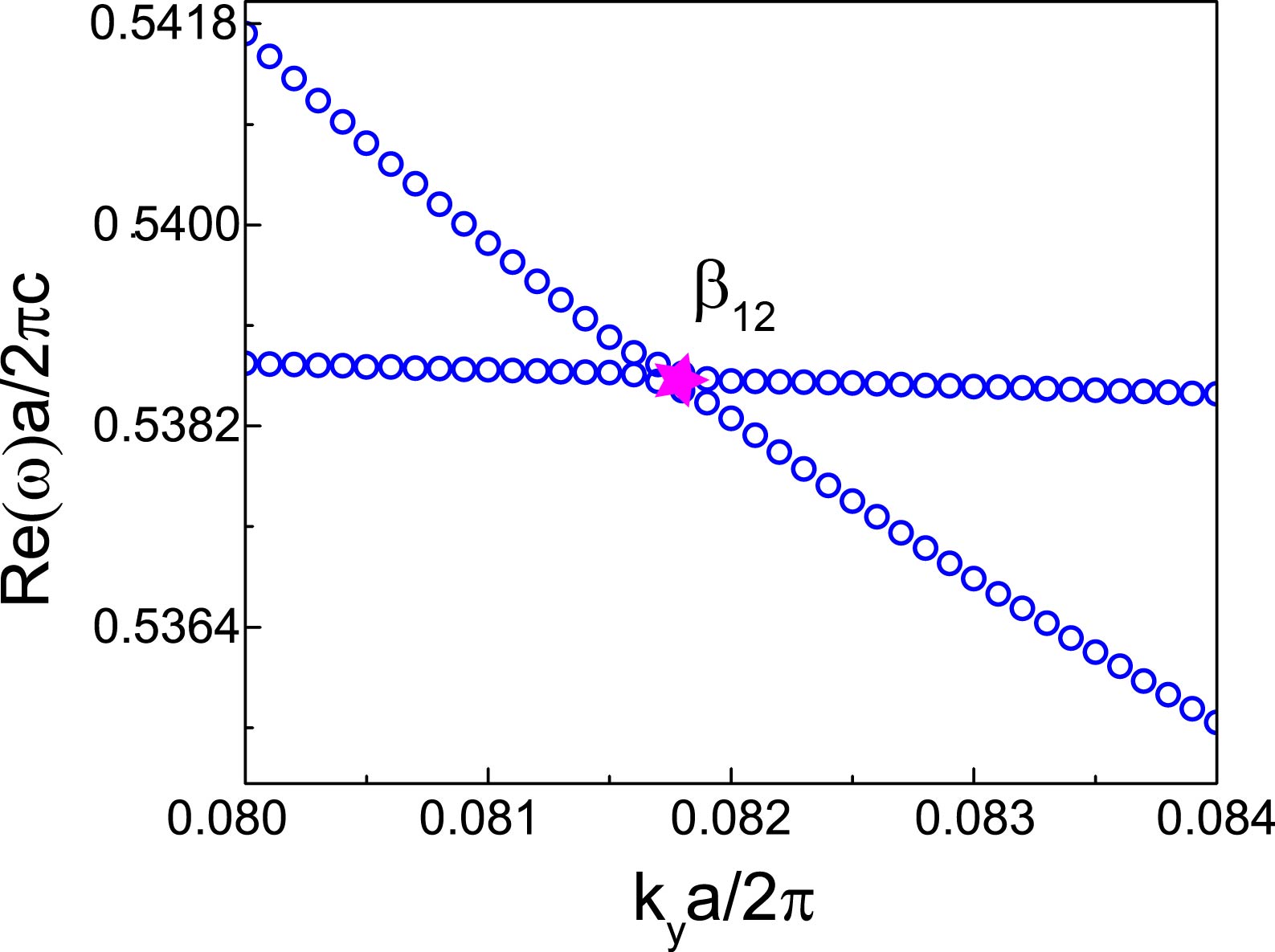}
	\caption{(Color online) Real parts of eigenfrequencies of interface states as functions of $ k_y $ are calculated by COMSOL with PML boundary conditions. The number of column layers of the semi-infinite PCs is truncated to $ N=15 $. The relative permittivity of the cylinders we used is $ \varepsilon_{c}=12 \pm 2.07706 i $. The pink star labels the anisotropic EP $ \beta_{12} $. }
	\label{fig:anisotropicbeta12ky}
\end{figure}

Note that in Figs. \ref{fig:bandinversionbeta} and \ref{fig:bandinversionbetaPi}, we have described the transition from EP $ S_{23}^L $ to EP $ S_{12}^L $ as $ \varepsilon_r $ increases. In Fig. \ref{fig:thirdorderandanisotripicEPs}, we plot the transition from EP $ S_{23}^L $ to EP $ S_{12}^L $ as $ \gamma $ increases and find the middle band touches the lowest band, forming an EP denoted by $ \beta_{12} $, which is shown in Fig. \ref{fig:anisotropicbeta12ky}.  
As $ \gamma $ increases further, the EP $ \beta_{12} $ splits into two EPs labeled as $ S_{12}^N $ and $ S_{12}^L $, which is shown in Fig.  \ref{fig:thirdorderandanisotripicEPs}(b). In Figs. \ref{fig:anisotropicbetaalpha}(d-f), we plot the EP $ \beta_{12} $ in the $ \gamma $ direction for a fixed $ k_y a/2 \pi = 0.0818 $. 
As shown in Figs. \ref{fig:anisotropicbeta12ky}  and  \ref{fig:anisotropicbetaalpha}(d-f), $\beta_{12} $ is an anisotropic EP, whose dispersion is linear along the $ k_y $ direction and of square-root form along the $ \gamma $ direction. We note that in the green line, two anisotropic EPs (denoted by $ \alpha_{12} $ and $ \beta_{12} $) appears. Different from EP $ \alpha_{12} $, whose imaginary parts of the eigenfrequency show linear dispersion along $ k_y $ direction, the real parts show linear dispersion around the EP $ \beta_{12} $.

\section{\label{sec:7}Conclusion}

In this work, we studied the formation of interface modes in $\mathcal{PT}$-symmetric PCs. It is well known that interface modes will exist if the real part of permittivity changes sign across the boundary of a PC, and we see here that interface modes with real eigenfrequencies will also exist if the imaginary parts of permittivity changes sign across a boundary. Best illustrated in Fig. \ref{fig:interfacebandpattern}, which shows the boundary mode dispersions for various values of the real part and imaginary part of the permittivity, the interface modes show some peculiar dispersion features that are not found in Hermitian systems. These interesting features include ziz-zag dispersions with turning points being exceptional points and there are also closed-loops of boundary modes with the vertices being exceptional points.
The peculiar band diagrams are quite similar to the folded bands with infinite group velocity points discussed previously\cite{Davanco2007,Chen2011}. These infinite group velocity points in the bands can be treated as exceptional points. We note in passing that a physical system carrying folded band has to be non-Hermitian. Hermitian systems (for example, those with $ \varepsilon $ equal to a constant negative value) can exhibit folded bands but those systems are not compatible with causality.
The trajectories of the EPs in the parameter space $ (k_y, \gamma) $ for different values of $ \varepsilon_r $ are summarized in Fig. \ref{fig:EPsinkgamma}. Close to the Brillouin zone center, when the Bloch momentum $ k_y \to 0 $, the existence and the dispersion of these boundary modes can be obtained semi-analytically using effective medium theory. This is illustrated in Fig. \ref{fig:EMTandBandinversion} and elaborated in the associated discussion. As the magnitude of Bloch momentum of the boundary modes increases, a more elaborate Hamiltonian model gives a quantitative description of the dispersion, which shows that the dispersion and its dependence on system parameters are closely related to the formation and coalescence of EPs. The details are rather fascinating, as we observe the coalescence of order-2 EPs into higher order EPs and the formation of anisotropic EPs. The results obtained using different computational methods (COSMOL and multiple scattering) and different boundary conditions are consistent with each other. The results obtained with periodic boundary conditions, as shown in Fig. \ref{fig:interface_periodic_x_projectband}, deserves some additional comments. First, it shows that as the size of the supercell increases, the bulk modes acquire an imaginary part in the eigenfrequency (entering the broken phase). In $\mathcal{PT}$-symmetric system, it is usually the increase in non-Hermiticity (“$ \gamma $”) that drives the system into the broken phase. The results in Fig. \ref{fig:interface_band_diagram} shows that the increase of system size (the number of cylinders “$ N $”) alone can take the system into the broken phase while $ \gamma $ remains constant. In the large $ N $ limit, only the interface modes remain in the exact phase, having zero imaginary part in its eigenfrequency and this band is localized on the loss-gain interface. This mode lies inside the continuum of the real bulk modes when $ \gamma=0 $, and in a certain sense these boundary modes are bound states in the continuum.

%\nocite{*}

\begin{acknowledgments}
This work is supported by the Research Grants Council, University Grants Committee, Hong Kong, through grants No. AoE/P-02/12, N\_HKUST608/17 and C6013-18G and by the National Natural Science Foundation of China (Grant Nos. 11761161002). K.D. acknowledges funding from the Gordon and Betty Moore Foundation. We would like to thank Prof. Zhao Qing Zhang, Prof. Shubo Wang and Dr. Ruoyang Zhang for helpful discussions.
\end{acknowledgments}

\appendix

\section{\label{sec:appA}Multiple scattering method}

In this appendix, we will construct the plane wave scattering problem of a 2D PC. We can determine the band structure of a PC from the solution of an eigenvalue problem \cite{Botten2001}. We consider a 2D PC comprising a rectangular lattice (lattice constant is $ a $ along $ y $-direction and $ b $ along $ x $-direction) of dielectric cylinders embedded in the air whose relative permittivity is equal to 1. The cylinders have a radius \(r_c\) , relative permittivity \(\varepsilon_c\), and relative permeability \(\mu_c=1\). 

We begin with a single infinite grating periodically arranged along \(y\)-direction, corresponding to one layer of this lattice. A TM polarization (electric field is along the axes of the cylinders) plane wave with wave vector \(\textbf{k}_0\) is incident onto the grating and perpendicular to the axes of the cylinders (\(z\)-direction). The wave number is \({k}_0^2=(\omega/c)^2\), where \(\omega\) is the angular frequency and \(c\) is the speed of light in vacuum. The component parallel to \(y\)-direction is denoted by \(k_p\), and therefore, the wave vector along $x$ direction is $k_x=\sqrt{k_0^2-k_p^2}$. We define the incident electric field as
\begin{equation}
    {\textbf{E}}^{\mathrm{inc}}=E_{0} \exp \left[i\left(k_{p} y+k_{x} x\right)\right] \hat{z},
    \label{eq:appB_Ein}
\end{equation}
where $E_0$ is the amplitude of the incident field.
Then the incident magnetic field can be written as
\begin{subequations}
\begin{eqnarray} H_{x}^{\mathrm{inc}} &=\frac{k_{p}}{\omega \mu_{0}} E_{0} \exp \left[i\left(k_{p} y+k_{x} x\right)\right] \\ H_{y}^{\mathrm{inc}} &=-\frac{k_{x}}{\omega \mu_{0}} E_{0} \exp \left[i\left(k_{p} y+k_{x} x\right)\right] \end{eqnarray}
\label{eq:appB_Hin}
\end{subequations}

To calculate the band structure, we need firstly establish the scattering matrices of the structure. The scattering field of the cylinders can be expanded as a summation of cylindrical functions, and then we expand these cylindrical functions as summation of plane waves at the interface of the grating. The various scattering plane waves with wave vector in the $x-y$ plane are like below,
\begin{equation}
        {{\bf{K}}_m ^ \pm } =  \left( {\pm q_m ,{k_p} - {g_m}} \right),
\label{eq:appendixB_Ksca}
\end{equation}
where $g_m=2m \pi/a$ is the reciprocal vector parallel to the grating, $q_m= \sqrt {{k_0}^2 - |{k_p} - {g_m}{|^2}} $ and the sign ($\pm$) before $q_m$ is corresponding to a propagating (if $q_m$ is real) or evanescent wave (if $q_m$ is imaginary) along the positive or negative $x$-direction. 

The full scattering matrices of the grating composed of cylinder are well established in many references, such as \cite{Botten2000}. These scattering matrices are usually complicated to calculate. For a 2D PC with TM polarization, we can represent the cylinders using an out-of-plane electric monopole and two in-plane magnetic dipoles \cite{Wang2016} 
\begin{subequations}
\begin{eqnarray}
{P}\hat z &= {\alpha ^e}E_z^{loc}\hat z ,\\
{\bf{M}} &= {\alpha ^m}{\bf{H}}_{}^{loc}.
\end{eqnarray}
    \label{eq:appB_dipoleandmonopole}
\end{subequations}
$E_z^{\text{loc}}$ and ${\textbf{H}}^{\text{loc}}$ are the local fields at the position of the cylinders. The local fields are summations of the external incident fields and the scattering field from other cylinders. $\alpha ^e$ and $\alpha^ m$ are the polarizability of the monopole and dipole of a cylinder, respectively, which are represented as \cite{Bohren1998}
\begin{subequations}
    \begin{eqnarray} \alpha^{e} &=\frac{4 i \varepsilon_{0}}{k_{0}^{2}} \beta_{0}\left(r_{c} k_{0}\right), \\ \alpha^{m} &=\frac{8 i}{k_{0}^{2}} \beta_{1}\left(r_{c} k_{0}\right), \end{eqnarray}
\label{eq:appB_polarizability}
\end{subequations}
where $\beta_0$ and $\beta_1$ are Mie scattering coefficients and can be represented using Bessel functions like below
\begin{widetext}
\begin{equation}
    {\beta _n}\left( {{r_c}{k_0}} \right) = \frac{{{\mu _c}{J_n}\left( {{n_c}{r_c}{k_0}} \right)J{'_n}\left( {{r_c}{k_0}} \right) - {n_c}{J_n}\left( {{r_c}{k_0}} \right)J{'_n}\left( {{n_c}{r_c}{k_0}} \right)}}{{{\mu _c}{J_n}\left( {{n_c}{r_c}{k_0}} \right)H{'_n}^{(1)}\left( {{r_c}{k_0}} \right) - {n_c}H_n^{\left( 1 \right)}\left( {{r_c}{k_0}} \right)J{'_n}\left( {{n_c}{r_c}{k_0}} \right)}}.
\label{eq:appB_Miescatt}
\end{equation}
\end{widetext}
$J_n$, $H_n^{(1)}$, ${J'}_n$, ${H'}_n^{(1)}$ are the first kind of Bessel functions, the first kind of Hankel functions and their derivatives with order $n$. $n_c=\sqrt{\varepsilon_c \mu_c}$ is the refractive index of the cylinder and $r_c$ is the radius of the cylinder. At the limit where $k_0 r_c$ is a small number, the scattering properties of the cylinder are well represented by the lowest three modes.  The electric fields induced by electric moment $P\hat z$ and magnetic moment $\textbf{M}$ are represented using Green’s function like below,
\begin{equation}
    \begin{aligned}
{\bf{E}}({\bf{r}})
 = \left[ {{P}{\omega ^2}\mu \hat z +i \omega \mu \nabla  \times {\bf{M}}} \right]\mathop {\bf{G}}\limits^ \leftrightarrow  \left( {{\bf{r}},{\bf{r'}}} \right)
\end{aligned}
\label{eq:appB_EscuingG}
\end{equation}
$\mathop {\bf{G}}\limits^ \leftrightarrow  \left( {{\bf{r}},{\bf{r'}}} \right)$ is a dyadic Green's function, representing the response at \textbf{r} induced by a point source at \(\textbf{r}'\).

Now let us turn to the PC illustrated in Fig. \ref{fig:interface_band_diagram}(a). We denote the field acting on the polarizable unit locating at $i$th column and $j$th row as ${E}_{ij}^{\text{loc}}\hat z$ and $\textbf{H}_{ij}^{\text{loc}}$ , $i = 1, 2,\cdots, N_1+N_2$, with $N_1$, $N_2$ being the number of column layers of the left and right sublattices. According to Eq. \eqref{eq:appB_dipoleandmonopole}, the induced monopole/dipole moments are ${{\bf{P}}_{ij}} = \alpha _i^e{{\bf{E}}_{ij}^{\rm loc}}$, ${{\bf{M}}_{ij}} = \alpha _i^m{{\bf{H}}_{ij}^{\rm loc}}$. The polarizability $\alpha_i^e$, $\alpha_i^m$ representing the cylinders in different columns could be different. The PC is uniform along $z$ direction, and therefore the fields are also uniform along $z$ direction. The electric monopole moment of the cylinder is polarized along $z$ direction, and the scattered electric fields of the cylinders only have $z$ component. The dyadic green function in Eq. \eqref{eq:appB_EscuingG} just keep $G_{zz}$ component as below
% \href{http://www.eecs.umich.edu/courses/eecs730/lect/DyadicGF_W09_port.pdf}{[dyadic Green's function]}
\begin{equation}
    G_{zz} (\textbf{r},\textbf{r}')=i/4H_0^{(1)} ( k_0 |\textbf{r}-\textbf{r}'|).
\end{equation}
Then the scattering electric fields  induced by a cylinder at $\textbf{r}=\textbf{r}_{ij}$ can be written as
\begin{equation}
   E^{\rm ind}_{ij}(\textbf{r}) \hat{z}= \left[ {{P_{ij}}{\omega ^2}\mu \hat{z}  +i \omega \mu \hat{z}\cdot \nabla  \times {\textbf{M}_{ij}}} \right] G_{zz} \left( {{\bf{r}},{\textbf{r}_{ij}}} \right).
\label{eq:appB_Escaij}
\end{equation}
The magnetic field can be obtained by Maxewell's equation like below
\begin{widetext}
\begin{equation}
   {\bf{H}}^{\rm ind}_{ij}({\bf{r}}) = \frac{1}{{i\omega \mu }}\nabla  \times E^{\text{ind}}\left( {\bf{r}} \right)\hat z =  - i\omega \nabla  \times \left[ {{P_{ij}}{G_{zz}}\left( {{\bf{r}},{{\bf{r}}_{ij}}} \right)\hat z} \right] + \nabla  \times \left[ {\hat z \cdot \nabla  \times {{\bf{M}}_{ij}}{G_{zz}}\left( {{\bf{r}},{{\bf{r}}_{ij}}} \right)} \right].
\label{eq:appB_Hscaij}
\end{equation}
%\end{widetext}
We know that the system is periodic along $y$-direction, applying the Bloch condition along $y$-direction into the polarization, 
\begin{equation}
     P_{ij} =\exp \left(i a_j k_{p}\right) P_{i}, \qquad \qquad {\bf{M}}_{ij} =\exp \left(i a_j k_{p}\right) {\bf{M}}_{i}, 
\label{eq:appB_Blochcon}
\end{equation}
where $P_{i}$ and $\textbf{M}_{i}$ represent the polarization of cylinder at $i$th column and $0$th row, and $a_j=r_{ij}-r_{i0}$ represents the distance between cylinders at $ r_{ij} $ and $ r_{i0} $.  
Then the total scattering field of the structure can be written as 
%\begin{widetext}
\begin{equation}
\begin{aligned}
{E^{{\rm{tot}}}}\left( {\bf{r}} \right) &= \sum\limits_{ij} {\left[ {{P_{ij}}{\omega ^2}\mu \hat z + i\omega \mu \hat z \cdot \nabla  \times {{\bf{M}}_{ij}}} \right]{{\rm{G}}_{zz}}\left( {{\bf{r}},{{\bf{r}}_{ij}}} \right)} \\
&= \sum\limits_i {\left[ {{P_i}{\omega ^2}\mu \hat z + i\omega \mu \hat z \cdot \nabla  \times {{\bf{M}}_i}} \right]\sum\limits_j {{{\rm{G}}_{zz}}\left( {{\bf{r}},{{\bf{r}}_{ij}}} \right)\exp \left( {i{a_j}{k_p}} \right)} } \\
&= \sum\limits_i {\left[ {{P_i}{\omega ^2}\mu \hat z + i\omega \mu \hat z \cdot \nabla  \times {{\bf{M}}_i}} \right]{\rm{G}}\left( {{\bf{r}},{{\bf{r}}_i}} \right)} ,
\end{aligned}
\end{equation}
%\end{widetext}
where we have defined 
\begin{equation}\label{eq:Gexpansion}
\begin{aligned}
    {\rm{G}}\left( {{\bf{r}},{{\bf{r}}_i}} \right)&=\sum\limits_j {{{\rm{G}}_{zz}}\left( {{\bf{r}},{{\bf{r}}_{ij}}} \right)\exp \left( {i{a_j}{k_p}} \right)}\\
    &=\frac{i}{4}\sum\limits_j {H_0^{\left( 1 \right)}\left( {{k_0}|{\bf{r}} - {{\bf{r}}_{ij}}|} \right)\exp \left( {i{a_j}{k_p}} \right)} \\
    &=\sum\limits_m^{} {\frac{{ - 1}}{{2iaq_m}}} \exp \left[ {i\left( {{k_p} - {g_m}} \right)y + iq_m |x - {x_{i0}}|} \right]
\end{aligned}
\end{equation}
representing the Green's function of the $i$th column and $ x_{i0} $ is the $ x $ coordinate of the cylinder at 
${{\bf{r}}_{i0}} = \left( {{x_{i0}},0} \right) $. In the last step of Eq. \eqref{eq:Gexpansion}, we use the identity for Hankel function to transform the lattice sum in real space into the reciprocal space \cite{Gomez-Medina2006,Botten2000}.

The local field acting on the cylinder at the position $\textbf{r}_{ij}$ is given by the summation of the external incident field and the scattering field induced by other cylinders at $\textbf{r} \ne \textbf{r}_{ij}$, that is, 
\begin{equation}
\begin{aligned}
    E_{ij}^{\text{loc}}&=E_{ij}^{\text{inc}}+{\left[ {{P_i}{\omega ^2}\mu \hat z + i\omega \mu \hat z \cdot \nabla  \times {{\bf{M}}_i}} \right]{\rm{G}_b}\left( {{{\bf{r}}_i}} \right)} +\sum\limits_{i' \ne i} {\left[ {{P_{i'}}{\omega ^2}\mu \hat z + i\omega \mu \hat z \cdot \nabla  \times {{\bf{M}}_{i'}}} \right]{\rm{G}}\left( {{\textbf{r}_{ij}},{{\bf{r}}_{i'}}} \right)}   
\end{aligned}
\label{eq:appB_Elocij}
\end{equation}
where 
%\begin{widetext}
\begin{eqnarray}
{G_b}\left( {{{\bf{r}}_i}} \right) &= \mathop {\lim }\limits_{{\bf{r}} \to {{\bf{r}}_{ij}}} G\left( {{\bf{r}},{{\bf{r}}_i}} \right) - {G_{zz}}\left( {{\bf{r}},{{\bf{r}}_{ij}}} \right) = \mathop {\lim }\limits_{{\bf{r}} \to {{\bf{r}}_{ij}}} \sum\limits_{j' \ne j} {{{\rm{G}}_{zz}}\left( {{\bf{r}},{{\bf{r}}_{ij'}}} \right)\exp \left( {i{a_{j'}}{k_p}} \right)}.
\label{eq:appB_Gb}
\end{eqnarray}
%\end{widetext}
The first term in Eq. \eqref{eq:appB_Elocij} represents the incident field, the second term represents the scattering fields induced by the column $i$, and the last term represents the scattering fields of other columns at $i' \ne i$. 
$G_b$ in Eq. \eqref{eq:appB_Gb} represents the total scattering field at $\textbf{r}_{ij}$ of $i$th column except for the cylinder at $\textbf{r}_{ij}$, which is expressed as \cite{Gomez-Medina2006} 
%\begin{widetext}
\begin{equation}
G_b=\frac{1}{{2\pi }}\left[ {\ln \left( {\frac{{{k_0}a}}{{4\pi }}} \right) + {\gamma _E}} \right] + i\left( {\frac{1}{{2ak_x }} - \frac{1}{4}} \right) + \frac{1}{{2a}}\sum\limits_{m = 1}^\infty  {\left( {\frac{i}{{q_m }} + \frac{i}{{q_{-m}}} - \frac{2}{{{g_m}}}} \right)}
\end{equation}
where $ \gamma_E $ is the Euler constant.

Substituting $ {P_i} = \alpha _i^eE_i^{{\rm{loc}}} $ and ${{\bf{M}}_i} = \alpha _i^m{\bf{H}}_i^{{\rm{loc}}}$ into Eq. \eqref{eq:appB_Elocij}, we can build the self-consistent equations like below
\begin{equation}
\sum\limits_{i' = 1}^{{N_1} + {N_2}} {\left[ {\begin{array}{*{20}{c}}
		{{\delta _{ii'}} - {\mu _0}{\omega ^2}\alpha _{i'}^{\rm{e}}{F_1}\left( {i,i'} \right)}&{i\omega {\mu _0}\alpha _{i'}^{\rm{m}}{F_3}\left( {i,i'} \right)}&{ - i\omega {\mu _0}\alpha _{i'}^{\rm{m}}{F_2}\left( {i,i'} \right)}\\
		{i\omega {\mu _0}\alpha _{i'}^{\rm{e}}F_3{{\left( {i,i'} \right)}}}&{{\delta _{ii'}} + \alpha _{i'}^{\rm{m}}{F_6}\left( {i,i'} \right)}&{ - \alpha _{i'}^{\rm{m}}{F_5}\left( {i,i'} \right)}\\
		{ - i\omega \alpha _{i'}^{\rm{e}}{F_2}\left( {i,i'} \right)}&{ - \alpha _{i'}^{\rm{m}}{F_5}\left( {i,i'} \right)}&{{\delta _{ii'}} + \alpha _{i'}^{\rm{m}}{F_4}\left( {i,i'} \right)}
		\end{array}} \right]}  \times \left[ {\begin{array}{*{20}{c}}
	{E_{i'}^{{\rm{loc}},z}}\\
	{H_{i'}^{{\rm{loc}},x}}\\
	{H_{i'}^{{\rm{loc}},y}}
	\end{array}} \right] = \left[ {\begin{array}{*{20}{c}}
	{E_{i'}^{{\rm{inc}},z}}\\
	{H_{i'}^{{\rm{inc}},x}}\\
	{H_{i'}^{{\rm{inc}},y}}
	\end{array}} \right]
\label{eq:couplediopleequation}
\end{equation}
%\begin{equation}
%\sum\limits_{i' = 1}^{{N_1} + {N_2}} {\left[ {\begin{array}{*{20}{c}}
%		{{\delta _{ii'}} - {\mu _0}{\omega ^2}\alpha _{i'}^{\rm{e}}{F_1}}&{i\omega {\mu _0}\alpha _{i'}^{\rm{m}}{F_3}}&{ - i\omega {\mu _0}\alpha _{i'}^{\rm{m}}{F_2}}\\
%		{i\omega {\mu _0}\alpha _{i'}^{\rm{e}}{F_3}}&{{\delta _{ii'}} + \alpha _{i'}^{\rm{m}}{F_6}}&{ - \alpha _{i'}^{\rm{m}}{F_5}}\\
%		{ - i\omega \alpha _{i'}^{\rm{e}}{F_2}}&{ - \alpha _{i'}^{\rm{m}}{F_5}}&{{\delta _{ii'}} + \alpha _{i'}^{\rm{m}}{F_4}}
%		\end{array}} \right]}  \times \left[ {\begin{array}{*{20}{c}}
%	{E_{i'}^{\text{loc},z}}\\
%	{H_{i'}^{\text{loc},x}}\\
%	{H_{i'}^{\text{loc},y}}
%	\end{array}} \right] = \left[ {\begin{array}{*{20}{c}}
%	{E_{i'}^{\text{inc},z}}\\
%	{H_{i'}^{\text{inc},x}}\\
%	{H_{i'}^{\text{inc},y}}
%	\end{array}} \right]
%\label{eq:couplediopleequation}
%\end{equation}
$ F_n (i,i')\quad (n=1, 2 \cdots 6)$ represents the lattice sums of the cylinder in column $ i' $ along the $ y $-direction and take expressions as below,

\[\begin{array}{l}
{F_1}\left( {l,{l^\prime }} \right) = \left\{ {\begin{array}{*{20}{l}}
	{\frac{1}{{2\pi }}\left[ {\ln \left( {\frac{{{k_0}a}}{{4\pi }}} \right) + {\gamma _E}} \right] + i\left( {\frac{1}{{2a{k_x}}} - \frac{1}{4}} \right) + \frac{1}{{2a}}\sum\limits_{m = 1}^\infty  {\left( {\frac{i}{{{q_m}}} + \frac{i}{{{q_{ - m}}}} - \frac{2}{{{g_m}}}} \right)} }&{l = {l^\prime },}\\
	{\sum\limits_m {\frac{i}{{2a{q_m}}}} {e^{i\left| {l - {l^\prime }} \right|b{q_m}}}}&{l \ne {l^\prime },}
	\end{array}} \right.\\
\\
{F_2}\left( {l,{l^\prime }} \right) = \left\{ {\begin{array}{*{20}{l}}
	0&{l = {l^\prime },}\\
	{{\mathop{\rm sign}\nolimits} \left( {l - {l^\prime }} \right)\sum\limits_m {\frac{{ - 1}}{{2a}}} {e^{i|l - {l^\prime }|b{q_m}}}}&{l \ne {l^\prime },}
	\end{array}} \right.\\
\\
{F_3}\left( {l,{l^\prime }} \right) = \left\{ {\begin{array}{*{20}{l}}
	{ - \frac{{{k_p}}}{{2a{k_x}}} - i\frac{{{k_p}}}{{2\pi }} - \sum\nolimits_{m = 1}^\infty  {\left[ {\frac{{{k_p} - {g_m}}}{{2a{q_m}}} + \frac{{{k_p} + {g_m}}}{{2a{q_{ - m}}}}} \right]} }&{l = {l^\prime },}\\
	{\sum\limits_m {\frac{{{g_m} - {k_p}}}{{2a{q_m}}}} {e^{i\left| {l - {l^\prime }} \right|b{q_m}}}}&{l \ne {l^\prime },}
	\end{array}} \right.\\
\\
{F_4}\left( {l,{l^\prime }} \right) = \left\{ {\begin{array}{*{20}{l}}
	{ - \frac{{{k_0}^2}}{{4\pi }}\left[ {\ln \left( {\frac{{{k_0}a}}{{4\pi }}} \right) + {\gamma _E} - \frac{1}{2}} \right] - \frac{{k_p^2}}{{4\pi }} - \frac{\pi }{{6{a^2}}} + i\left( {\frac{{k_0^2}}{8} - \frac{{{k_x}}}{{2a}}} \right) - \frac{1}{{2a}}\sum\limits_{m = 1}^\infty  {\left( {i{q_m} + i{q_{ - m}} + 2{g_m} - \frac{{k_0^2}}{{{g_m}}}} \right)} }&{l = {l^\prime },}\\
	{\sum\limits_m {\frac{{ - i{q_m}}}{{2a}}} {e^{i|l - l'|b{q_m}}}}&{l \ne {l^\prime },}
	\end{array}} \right.\\
\\
{F_5}\left( {l,{l^\prime }} \right) = \left\{ {\begin{array}{*{20}{l}}
	0&{l = {l^\prime }}\\
	{{\mathop{\rm sign}\nolimits} \left( {l - {l^\prime }} \right)\sum\limits_m {\frac{{i\left( {{g_m} - {k_p}} \right)}}{{2a}}} {e^{i|l - {l^\prime }|b{q_m}}}}&{l \ne {l^\prime }}
	\end{array}} \right.\\
\\
{F_6}\left( {l,{l^\prime }} \right) = \left\{ {\begin{array}{*{20}{l}}
	{ - \frac{{{k_0}^2}}{{4\pi }}\left[ {\ln \left( {\frac{{{k_0}a}}{{4\pi }}} \right) + {\gamma _E} + \frac{1}{2}} \right] + \frac{{k_p^2}}{{4\pi }} + \frac{\pi }{{6{a^2}}} + i\left( {\frac{{k_0^2}}{8} - \frac{{k_p^2}}{{2b{k_x}}}} \right) - \frac{1}{{2a}}\sum\limits_{m = 1}^\infty  {\left( {\frac{{i{{\left( {{k_p} - {g_m}} \right)}^2}}}{{{q_m}}} + \frac{{i{{\left( {{k_p} + {g_m}} \right)}^2}}}{{{q_{ - m}}}} - 2{g_m} - \frac{{k_0^2}}{{{g_m}}}} \right)} }&{l = {l^\prime },}\\
	{\sum\limits_m {\frac{{ - i\left( {{g_m} - {k_p}} \right)}}{{2a}}} {e^{i|l - l'|b{q_m}}}}&{l \ne {l^\prime }.}
	\end{array}} \right.
\end{array}\]

\end{widetext}
In the above equations, $m$ is the diffraction order, and the sum over $m$ is truncated to certain value $m_0$ in numerical calculation. $ l-l'$ denotes the distance between different column, and sign($l-l'$) is the sign function. $ a $ and $ b $ are the lattice constant along $ y $- and $ x $-direction, respectively. For a square lattice that we studied in this paper, we have $ a=b $. 

Equation \eqref{eq:couplediopleequation} describes a scattering problem and can be rewritten as below,
\begin{equation}
    {M}\left( \omega, k_p  \right){\Psi ^{\rm loc}} = {\Psi ^{\rm inc}}
\label{eq:appB_scatteringeigenvalueproblem}
\end{equation}
where we denote the content in the first bracket of Eq.\eqref{eq:couplediopleequation} as a matrix $M\left( \omega, k_p  \right)$, ${\Psi ^{\rm loc}}$ represents incident wave from outside and ${\Psi ^{\text{loc}}}$ represents local field incident on the cylinder.

\section{\label{sec:appB}The left eigenstate of a non-Hermitian PC with reciprocal media}

In analogy with the inner product of two wave functions in quantum mechanics\cite{Sakuraibooks}, we can define the inner product of two vector fields $ \textbf{F}(\textbf{r}) $ and $ \textbf{G}(\textbf{r}) $ as
\begin{equation}
	\left\langle {\bf{F}} \right.\left| {\bf{G}} \right\rangle  \equiv \int {{d^3}} {\bf{r}}{{\bf{F}}^*}({\bf{r}}) \cdot {\bf{G}}({\bf{r}})
\end{equation}
The Hermitian adjoint of an operator $\Theta$ is labeled as $\Theta^\dagger$, and is defined as \cite{joannopoulos2008}
\begin{equation}
\left\langle\Theta^{\dagger} \mathbf{F} | \mathbf{G}\right\rangle \equiv\langle\mathbf{F} | \Theta \mathbf{G}\rangle.
\end{equation}
We say the operator $\Theta$ is Hermitian if $\Theta^\dagger=\Theta$.

In PCs, the eigenvalue problem for electric field can be written as \cite{sakoda2004optical}
\begin{equation}
\mathcal{L} \mathbf{E}(\mathbf{r}) \equiv \varepsilon^{-1} \nabla \times \nabla \times \mathbf{E}(\mathbf{r})=\left(\frac{\omega}{c}\right)^{2}  \mathbf{E}(\mathbf{r}),
\end{equation}
where $\mathcal{L} \equiv \varepsilon^{-1} \nabla \times \nabla \times$. The Hermitian adjoint of the operator $\mathcal{L}$ is 
\begin{equation}
    \mathcal{L}^{\dagger}=\nabla \times \nabla \times\left(\varepsilon^{-1}\right)^{\dagger}=\varepsilon \mathcal{L}\left(\varepsilon^{-1}\right)^{\dagger}
\end{equation}
 For a reciprocal medium $ \varepsilon^T=\varepsilon $, we have
 \begin{equation}\label{eq:appC-LE}
 \mathcal{L}^{T}=\left(\mathcal{L}^{\dagger}\right)^{*}=\nabla \times \nabla \times\left(\varepsilon^{-1}\right)^{T}=\varepsilon \mathcal{L} \varepsilon^{-1}.
 \end{equation}

In a periodic system, Bloch state  ${\bf{E}}_{\bf{k}}^R\left( {\bf{r}} \right) = {e^{i{\bf{k}} \cdot {\bf{r}}}}{\bf{u}}_{\bf{k}}^R\left( {\bf{r}} \right) $ is a common eigenstate of $ \mathcal{L} $ and the lattice translation operator $ T_\mathbf{a} $ ,
\begin{subequations}\label{eq:appC-commonE}
	\begin{eqnarray}
		{\cal L}{\bf{E}}_{\bf{k}}^R &= {\left( {\frac{{\omega \left( {\bf{k}} \right)}}{c}} \right)^2}{\bf{E}}_{\bf{k}}^R,\\
		 {T_{\bf{a}}}{\bf{E}}_{\bf{k}}^R &= {e^{i{\bf{k}} \cdot {\bf{a}}}}{\bf{E}}_{\bf{k}}^R.
	\end{eqnarray}
\end{subequations}
For the non-Hermitian system, we define the corresponding left eigen state as
\begin{equation}\label{eq:appC-LE-lefteigenvector}
{{\cal L}^\dag }{\bf{E}}_{\bf{k}}^L = {\left( {\frac{{\omega {{\left( {\bf{k}} \right)}^*}}}{c}} \right)^2}{\bf{E}}_{\bf{k}}^L.
\end{equation}
Taking complex conjugate of Eq. \eqref{eq:appC-LE-lefteigenvector} and multiply $ \varepsilon^{-1} $ on both sides, we obtain
\begin{equation}\label{eq:appC-LE-epsilonleft}
{\varepsilon ^{ - 1}}{\left( {{{\cal L}^\dag }{\bf{E}}_{\bf{k}}^L} \right)^*} = {\cal L}{\varepsilon ^{ - 1}}{\left( {{\bf{E}}_{\bf{k}}^L} \right)^*} = {\varepsilon ^{ - 1}}{\left( {\frac{{\omega \left( {\bf{k}} \right)}}{c}} \right)^2}{\left( {{\bf{E}}_{\bf{k}}^L} \right)^*}.
\end{equation}
Therefore, $ {\varepsilon ^{ - 1}}{\left( {{\bf{E}}_{\bf{k}}^L} \right)^*} $ is also a right eigenstate of $ \mathcal{L}$ due to the reciprocity of the permittivity. 
Since $ \left[ {{{\cal L}^\dag },{T_{\bf{a}}}} \right] = 0 $, the left eigenstate $ \mathbf{E}_{\mathbf{k}}^{L} $  should be an eigenstate of $ T_\mathbf{a} $, i.e. $ T_{\mathrm{a}} \mathbf{E}_{\mathrm{k}}^{L}=\lambda(\mathbf{a}) \mathbf{E}_{\mathrm{k}}^{L} $. And using the normalization relation of the Bloch states,
\begin{equation}\label{eq:appC-LE-normalization}
1=\left\langle\mathbf{E}_{\mathbf{k}}^{L} | \mathbf{E}_{\mathbf{k}}^{R}\right\rangle=\left\langle\mathbf{E}_{\mathbf{k}}^{L}|T_{\mathbf{a}}^{\dagger} T_{\mathbf{a}}| \mathbf{E}_{\mathbf{k}}^{R}\right\rangle=\lambda(\mathbf{a})^{*} e^{i \mathbf{k} a},
\end{equation}
we can obtain $ \lambda(\mathbf{a})=e^{i \mathbf{k} \cdot \mathbf{a}} $, which means that $ \mathbf{E}_{\mathrm{k}}^{L} $ is a Bloch state at \textbf{k}. We can check that
\begin{eqnarray}\label{eq:appC-LE-TaEL}
%\begin{aligned}
{T_{\bf{a}}}{\varepsilon ^{ - 1}}{\left( {{\bf{E}}_{\bf{k}}^L\left( {\bf{r}} \right)} \right)^*} &=& {\varepsilon ^{ - 1}}{\bf{E}}_{\bf{k}}^L{\left( {{\bf{r}} + {\bf{a}}} \right)^*} \nonumber\\ &=& {\varepsilon ^{ - 1}}{\left( {{e^{i{\bf{k}} \cdot {\bf{a}}}}{\bf{E}}_{\bf{k}}^L\left( {\bf{r}} \right)} \right)^*} \nonumber\\ &=& {e^{ - i{\bf{k}} \cdot {\bf{a}}}}{\varepsilon ^{ - 1}}{\bf{E}}_{\bf{k}}^L{\left( {\bf{r}} \right)^*},
%\end{aligned}
\end{eqnarray}
\\
which means $ \varepsilon^{-1}\left(\mathbf{E}_{\mathrm{k}}^{L}\right)^{*}=\mathbf{E}_{-\mathrm{k}}^{R} $ is the right Bloch state at $ -\mathbf k $, and therefore,
\begin{equation}\label{eq:appC-LE-EkLstar}
\mathbf{E}_{\mathbf{k}}^{L}=\left(\varepsilon \mathbf{E}_{-\mathbf{k}}^{R}\right)^{*}.
\end{equation}
The biorthonormal relationship of a non-Hermitian PC becomes 
\begin{eqnarray}
%\begin{aligned}
\left\langle {{\bf{E}}_{{\bf{k}}m}^L} \right.\left| {{\bf{E}}_{{\bf{k}}n}^R} \right\rangle  & = & \int {d^2}r{{\left( {{\bf{E}}_{{\bf{k}}m}^L} \right)}^*}{\bf{E}}_{{\bf{k}}n}^R \nonumber \\
& = &\int {d^2}r{\bf{E}}_{ - {\bf{k}}m}^R\varepsilon {\bf{E}}_{{\bf{k}}n}^R \nonumber\\
& = &\int {{d^2}r{\bf{u}}_{ - {\bf{k}}m}^R\varepsilon {\bf{u}}_{{\bf{k}}n}^R  }\nonumber \\&=& {\delta _{mn}}.
%\end{aligned}
\end{eqnarray}

\bibliography{refinter}% Produces the bibliography via BibTeX.

\end{document}